\documentclass[useAMS,usenatbib,a4paper]{mn2e}
\usepackage{times}
\title[Giant Molecular clouds]
{
The exciting lives of giant molecular clouds
}
\author[Dobbs]
{C. L. Dobbs\thanks{E-mail:
dobbs@astro.ex.ac.uk}$^{1}$ and J. E. Pringle$^{2}$\\
$^1$ School of Physics, University of Exeter, Stocker Road, Exeter, EX4 4QL, UK \\
$^2$ Institute of Astronomy, Madingley Road, Cambridge, CB3 0HA }
\usepackage{amssymb}
\usepackage{amsmath}
\usepackage{graphicx}
\usepackage{epsfig}
\usepackage{multirow}

\begin{document}
\label{firstpage}
\date{\today}

\pagerange{\pageref{firstpage}--\pageref{lastpage}} \pubyear{2012}

\maketitle

\begin{abstract}
We present a detailed study of the evolution of GMCs in a galactic disc simulation. We follow individual GMCs (defined in our simulations by a total column density criterion), including their level of star formation, from their formation to dispersal. We find the evolution of GMCs is highly complex. GMCs often form from a combination of smaller clouds and ambient ISM, and similarly disperse by splitting into a number of smaller clouds and ambient ISM. However some clouds emerge as the result of the disruption of a more massive GMC, rather than from the assembly of smaller clouds. Likewise in some cases, clouds accrete onto more massive clouds rather than disperse. Because of the  difficulty of determining a precursor or successor of a given GMC, determining GMC histories and lifetimes is highly non--trivial. Using a definition relating to the continuous evolution of a cloud, we obtain lifetimes typically of 4-25 Myr for $>10^5$ M$_{\odot}$ GMCs, over which time the star formation efficiency is about 1 \%. We also relate the lifetime of GMCs to their crossing time. We find that the crossing time is a reasonable measure of the actual lifetime of the cloud, although there is considerable scatter. The scatter is found to be unavoidable because of the complex and varied shapes and dynamics of the clouds. We study cloud dispersal in detail and find both stellar feedback and shear contribute to cloud disruption. We also demonstrate that GMCs do not behave as ridge clouds, rather massive spiral arm GMCs evolve into smaller clouds in inter-arm spurs.  
\end{abstract}


\section{Introduction}
Observational studies of molecular clouds are limited by providing only an instantaneous picture of where clouds lie and what their properties are. As such we have a restricted understanding of the evolution of molecular clouds, for example of how they form, what they form from, and for how long they undergo star formation. Numerical simulations offer one way to study cloud evolution. However even with simulations, studying the evolution of clouds is still challenging, with problems such as how to define a clump or cloud (see e.g. \citealt{Pineda2009,Reid2010,Hughes2013}), and because the constituent gas of clouds may change on relatively short timescales (e.g. Myrs, \citealt{Dobbs2011old}).

Observations of Galactic GMCs show that some, such as Orion, are actively forming stars, whereas others, such as California and Maddalena's cloud are much more quiescent \citep{Lada2009,Megeath2009}. Nevertheless all nearby GMCs appear to contain at least some star formation, seemingly since all contain some fraction of very dense ($\sim10^4$ cm$^{-3}$) gas \citep{Froebrich2010,Lada2010}. One implication of these observations, and the generally small age spread of stars in GMCs \citep{Efremov1998,Mayne2008,Sung2010,Jeffries2011}, is that the timescale for star formation is very short \citep{Larson1981,Elmegreen2000,Hart2001}. Likewise this implies a short  formation time for GMCs themselves \citep{Pringle2001}. To account for these short formation times, there have been two main possibilities - either diffuse gas is collected together in a colliding flow \citep{Audit2005,Heitsch2006,Vaz2007,Vaz2010,Inoue2012} on a Myr timescale, or gas is relatively dense prior to becoming a GMC \citep{Pringle2001}. In \citet{Dobbs2012} we considered the gas which goes to form the global population of GMCs at a given timeframe, and found that gas is typically overly dense for tens of Myr prior to GMC formation, supporting the latter scenario. Observing the  precursors of GMCs is difficult and it is not known whether they are dense clouds of HI, diffuse HI, or other molecular clouds. In M33 and the LMC, there have been observations of GMCs that do not contain massive star formation \citep{Kawamura2009,Gratier2012}, but there may well be unobserved low mass star formation in these clouds, similar to the California cloud.

Numerical simulations have also investigated the physical mechanisms for GMC formation \citep{DBP2006,Dobbs2008,Wada2008,Kim2002,Kim2003,Tasker2009}. In \citet{Dobbs2008} we showed that at low surface densities ($\lesssim 10$ M$_{\odot}$ pc$^{2}$) GMCs form from the collision or agglomeration of smaller clouds. As the surface density increases, gravitational attraction between the clouds enhances collisions \citep{Kwan1987} and gravitational instabilities in the gas play a more important role \citep{Kim2002,Kim2006}. This picture of agglomeration and self gravity is largely supported by recent observations of the GMCs and the properties of GMCs in M51 \citep{Colombo2013}.

The lifetimes of GMCs are not well known either observationally or theoretically (although as we point out in this paper the definition of a GMC lifetime is not clear either), although many observations point towards a lifetime of $\sim 20-30$ Myr. The first observational estimates based on the presence of CO surrounding clusters of different ages, suggested that the ages of Galactic GMCs are $\sim 20$ Myr \citep{Bash1977,Leisawitz1989,Blitz1993}. More recent results have estimated cloud lifetimes in the LMC and M33 to be 20-30 Myr \citep{Kawamura2009,Miura2012}. \citet{Murray2011} deduced similar lifetimes based on the estimated time for feedback to disrupt observed GMCs. There have however been suggestions of longer lifetimes, for example to explain the presence of inter-arm GMCs in the Milky Way \citep{Scoville1979} and other galaxies such as M51 \citep{Koda2009}. Over a cloud's lifetime some fraction of the cloud's mass it turned into stars, representing the star formation efficiency. This fraction is typically a few per cent and is thought to range from $\lesssim 1$\% to 20\% \citep{Evans2009,Murray2011}. The relatively low value is thought to be due to stellar feedback, which changes the dynamics and disrupts GMCs \citep{Dobbs2011new,Vaz2010} and/or magnetic fields \citep{Price2008,Padoan2011}.

The primary method of disruption of GMCs is also not well known. GMCs are subject to external processes, such as shear and turbulence, which can disperse the cloud, as well as internal processes of star formation and stellar feedback. Recently \citet{Dobbs2011old} argued that many or most GMCs in galaxies are unbound. For these, internal processes such as stellar feedback are not necessary to disrupt the cloud. For bound GMCs, and for bound substructures within clouds, disruption requires energy input from stellar feedback processes such as jets, winds, HII regions and supernovae.

In a previous paper \citep{Dobbs2012} we investigated the gas density of material before and after it forms molecular clouds. We found that gas is likely to be overly dense compared to the average densities of the ISM, at least 30 Myr prior to GMC formation. In particular, gas which reaches moderate densities of $\sim10$ cm$^{-3}$ persists for tens of Myr, even though dense ($\sim100$ cm$^{-3}$) gas is only very short-lived. Furthermore we found some examples where clouds were never completely dispersed into the ambient medium. Here we extend these ideas, but to show examples of individual clouds in detail. We follow their formation, dispersal and star formation history. We focus on a representative $10^6$ M$_{\odot}$ cloud, but also describe the evolution of other $10^5-10^6$ M$_{\odot}$ clouds, and show some overall statistics for $>10^5$ M$_{\odot}$ clouds. We also vary our criteria for defining a cloud to check the reliability of our results. 

\section{Calculations}
We performed a calculation with a 2 armed spiral potential, heating and cooling, self gravity and stellar feedback. The implementation of all these processes has been described in previous papers. The potential used comprises a logarithmic component \citep{Binney1987} and a spiral component \citep{Cox2002}, and is the same as that given in \citet{DBP2006}, except in this paper we choose $N=2$ rather than $N=4$. We changed to $N=2$ simply because for many roughly symmetric galaxies, the number of arms is 2 rather than 4 (the choice of the latter previously was also influenced partly by the supposed structure of our Galaxy (see \citealt{Cox2002} and \citealt{Vallee2005})).

The cooling and heating of the ISM we use is taken from \citet{Glover2007} and described more fully in \citet{DGCK2008}. Cooling is switched off if the temperature of a particle drops below 50 K. Also as in these previous papers we include time-dependent H$_2$ formation, described briefly as follows. Each particle is assigned a fraction of H$_2$ (from 0 to 1) and the rate of change of $n(H_2)$ is calculated according to the prescription given in the appendix of \citet{Bergin2004}, and \citet{DBP2006}.  This prescription includes the rate of formation of H$_2$ on grains (see e.g. \citealt{Holl1971b,Jura1974}) and a destructive photodissociation term \citep{Draine1996}. For the photodissociation term, we used a simple approximation of estimating the column density from the local density multiplied by a scale-length of 35 pc, as described in \citet{DGCK2008}.

We implement stellar feedback as described in \citet{Dobbs2011new}, but also provide the main details here. Although some recent work has considered in detail different mechanisms such as ionisation, winds and supernovae \citep{Hopkins2011,Agertz2012,Kim2012}, we simply insert an amount of energy where star formation occurs. We insert an amount of energy according to 
\begin{equation}
E=\frac{\epsilon M(H_2)} {160} \times 10^{51} ergs
\end{equation}
where $\epsilon$ is a constant which we take to be 0.05, M(H$_2$) is the mass of molecular hydrogen in a region of gas (dependent on the smoothing length at a particular resolution),  the division by 160 comes from our choice of IMF, and we multiply by 10$^{51}$ ergs assuming this is the energy released from one supernovae. For our resolution, each feedback event typically corresponds to 1 or 2 supernovae. However we note that our scheme could count collectively for all stellar feedback. The energy associated with processes such as jets is much smaller, whilst that for stellar winds and radiation pressure are comparable to supernovae \citep{MacLow2004,Agertz2012}, hence the net amount will not be greatly different from the $10^{51}$ ergs we assume. We deposit the energy as half kinetic and half thermal energy. 

The simulation presented here uses 8 million particles, each particle having a mass of 312.5 M$_{\odot}$. We also present a 1 million particle simulation in the appendix as a resolution test. The mean surface density of the galaxy is 8 M$_{\odot}$ pc$^{-2}$.

\section{Results: Evolution of galaxy}
In Figure~1 we show the column density for our galactic disc at a time of 250 Myr. The galaxy exhibits two clear spiral arms in the central parts, though the arms become weaker at larger radii. The difference in velocities across the shock drop substantially beyond $\sim$ 6 kpc reflecting that, although the depth of the potential (the spiral perturbation is of order 0.1 \% of the logarithmic component) does not alter significantly, the potential minimum is much broader and the gas does not undergo a strong shock.

The gas in the disc is clumpy, with the most massive, dense clumps or clouds situated in or near the spiral arms. These massive clouds form by a process of agglomeration of smaller clumps in the spiral arms, aided by self gravity. Clouds in the spiral arms are then sheared in to inter-arm spurs, producing a structure remarkably similar to observed grand design galaxies such as M51 and M74. From 150 Myr onwards, the gas lies in approximately one third of the cold ($<150$ K), intermediate and warm ($>7000$ K) phases of the ISM respectively. A small fraction of the gas, $\sim$ 1 \%, lies in the hot ($>10^6$ K) phase. We ran our calculation until a time of 300 Myr, and we carry out most of our analysis at, or around the time of 250 Myr.

\begin{figure}
\includegraphics[scale=0.62]{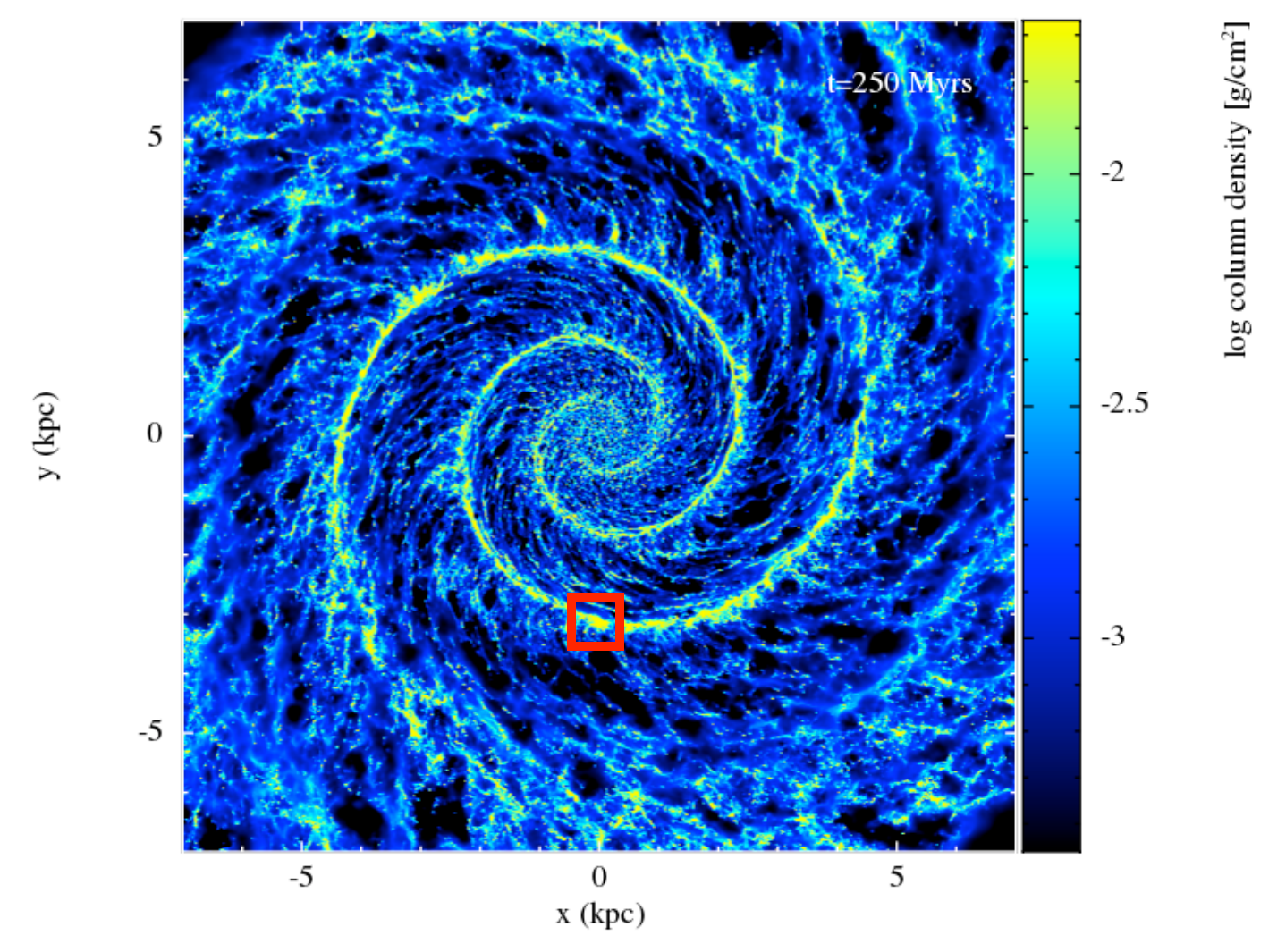}
\caption{The column density for the disc in our simulation is shown at a time of 250 Myr. The red box indicates the region shown in Figure 3 (middle panel) which contains Cloud380.}
\end{figure}

\section{Defining and tracing clouds}
In this section, we describe how we select clouds in  our simulation, how we trace them over time, and how we estimate the lifetime of a cloud. We demonstrate that the clouds do not behave as `ridge' clouds, so it appropriate to follow them as a collection of particles. We also discuss how cloud can be dispersed, both in terms of physical processes and with regards to the limitations of numerical simulations.

\subsection{Cloud selection}
We used a simple clump-finding algorithm which, for most of our analysis, selects clumps according to a surface density threshold criterion of $\Sigma_{th}=100$ M$_{\odot}$ pc$^{-2}$. The galaxy is divided into a Cartesian grid, with cells of size 10 pc, and all adjacent cells which exceed the density threshold are grouped together and classified as a cloud (cells which are only diagonally touching are not included). We also require that clouds contain at least 30 particles. At 250 Myr, this gave 1562 clouds with masses in the range $10^4$ M$_{\odot}$ to $2 \times 10^6$ M$_{\odot}$. We refer to any material which is not in a cloud as `ambient ISM'. In Section~5.3, and the appendix, we also show results where we vary the surface density threshold (referred to hereafter as $\Sigma_{th}$) used in our clump-finding algorithm.

\subsection{Properties of clouds}
In Appendix A, we show the mass spectrum, rotation and virial parameters of the clouds we detect. In addition, we also compare results with different surface density criteria, and results from a lower resolution simulation with only 1 million particles. Overall the properties we obtain are similar to the properties of clouds presented in \citet{Dobbs2011new}, although in that calculation we only used 1 million particles, and adopted a 4 arm spiral potential. We note also that the properties of GMCs in our main, 8 million particle simulation, extend to lower mass ($10^4$ M$_{\odot}$) clouds compared to \citet{Dobbs2011new}. 

\subsection{Tracing the evolution of clouds}
Since we are using a Lagrangian code, we are able to trace the constituent particles in a given cloud to earlier and to later times. However, when a cloud is defined at a particular moment in time (for example by using a clump-finding algorithm) it is only really defined at that instant. At the next instant (or at a previous instant) what appears to be the same cloud will in fact be a new cloud with (slightly) different constituent gas and different boundaries. At later times, the cloud may break up into one or more clouds -- some of the original cloud gas will be split among many other clouds, and some may have returned to the intercloud medium. Similarly, if we try to trace back the cloud to earlier times, the gas in the cloud will have come from a number of other clouds and/or directly from the intercloud medium. Thus the concept of the evolutionary history and lifetime of a cloud is not necessarily a particularly useful one. 

With this in mind, we set about tracing the evolution of a cloud in the following manner: (i) at a particular time $T_0$ (here chosen to be 250 Myr) we consider a particular cloud $A$ defined by the clump finding algorithm, and nominate the constituent particles of cloud $A$ at that time as `\textit{original cloud particles}', then (ii) at other times $t \neq T_0$ we use the same clump finding algorithm and list all those clouds which contain more than $20$ (i.e. $> 6250 M_\odot$) of the original cloud particles (Recall that by definition a cloud must contain more than 30 particles.). Then at each time $t$ we can identify that cloud which contains the greatest number of original cloud particles, and we nominate that cloud as the \textit{continuation} of cloud $A$ in time. 

In this way we are able to establish the evolution of a cloud in a Lagrangian sense, that is we define the evolving cloud in terms of the movement of a set of original cloud particles. 

\subsubsection{Molecular clouds are not `ridge clouds'}

As an aside here we note that this is not necessarily the only, nor the best, definition of cloud evolution, depending on circumstances. For example if the molecular cloud were the astronomical equivalent of a `ridge cloud' (that is, a rain cloud that forms near a mountain ridge as air is forced over it), perhaps being caused as gas is compressed by passing through a spiral arm, then the cloud would remain roughly in one place as its constituent particles passed through it. 

To show that this would not be a sensible procedure in this case we consider Cloud380 described in more detail in the next Section. At time $T_0$ we define the volume V which coincides with Cloud380. We then plot the mass of gas $M_V(t)$ which lies within this volume (in the co-rotating frame) in Figure~2, and compare this with the mass of original cloud particles $M_{V0}(t)$ within V at each time $t$. Were Cloud 380 a ridge cloud, then $M_V(t)$ would remain roughly constant, while the `original cloud particles' $M_{V0}(t)$ would flow through the volume. We can see from Figure~2 that $M_V$ and $M_{V0}$ track each other quite well. This suggests that Cloud 380 is not a ridge cloud, and that tracking its evolution by following its constituent particles  does make sense.

\begin{figure}
\includegraphics[scale=0.4]{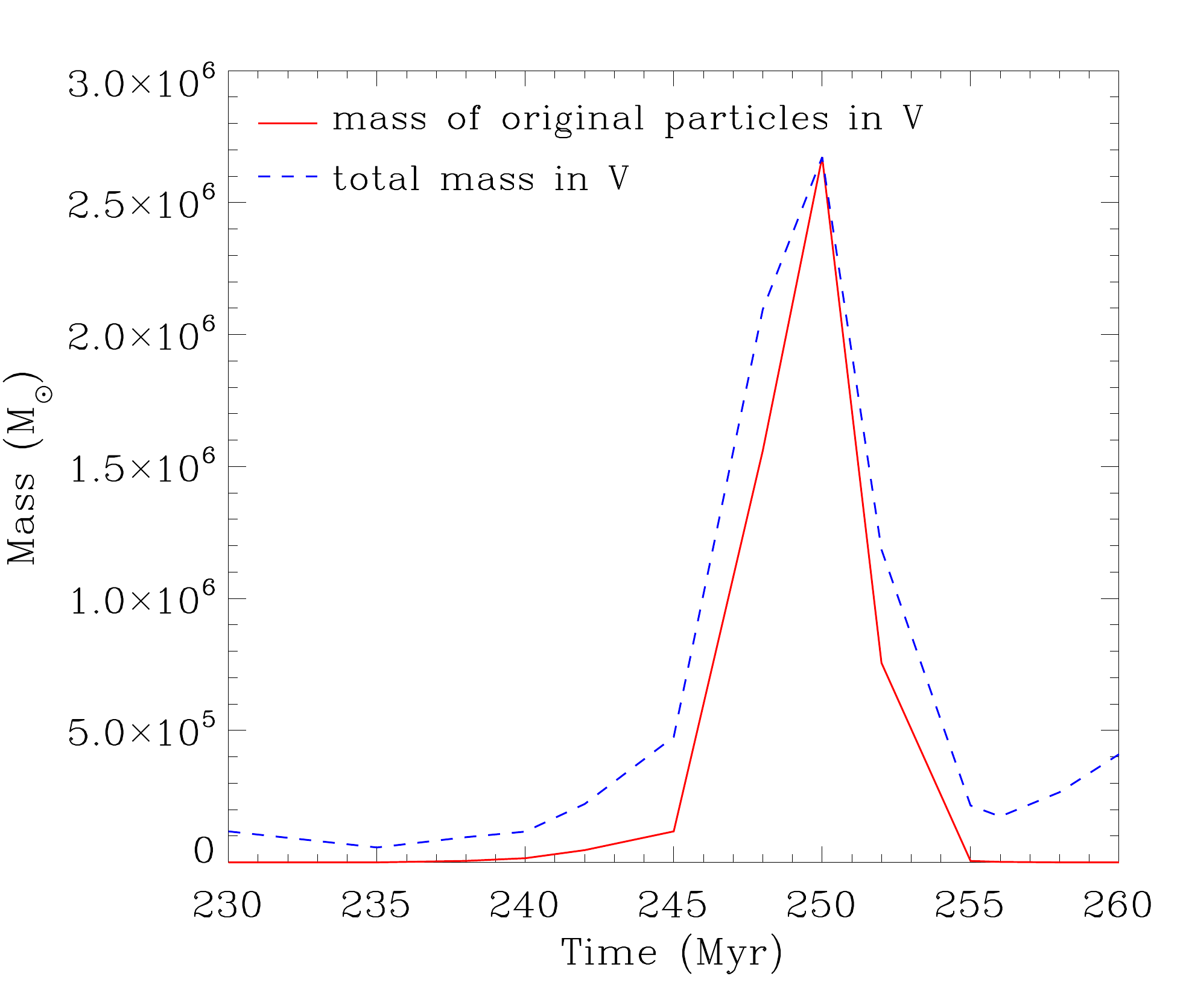}
\caption{The change in mass of gas in a volume $V$ is shown versus time. $V$ includes, and is slightly larger than Cloud380 (see Section 4.4). The two lines show the total mass of gas in $V$ versus time ($M_V(t)$ -- see text, blue dashed) and the mass of the original particles of $V$ at time $T_0=250$ Myr ($M_{V0}(t)$, red solid) versus time. Since the evolution of both is very similar, we conclude that cloud evolution is essentially Lagrangian and the cloud is not a ridge cloud.}
\end{figure}

\subsubsection{Dispersal of clouds}
As we show later, the dispersal of clouds in the simulations is highly complex. There are several physical mechanisms by which clouds can be dispersed, i) the simple unboundedness of the cloud, ii) shear, and iii) stellar feedback, or some combination of these. The complex structure of many clouds means it is easier to split clouds apart by these processes, e.g. if a cloud is a long filament, it is relatively easy to split the cloud into two if there is a feedback event, the cloud is subject to shear, or generally one part of the cloud is experiencing different large scale motions to the rest.   

There are also a couple of artificial ways that clouds can disperse (or form), due to the resolution of our simulation and method for finding clouds. A cloud can disappear completely simply by falling under the threshold used to determine clouds. For more massive clouds though it is unlikely that the whole cloud will disappear simultaneously this way. Also clouds form and disappear instantaneously if the number of particles changes between 29 and 30. As we restrict our studies to $>10^5$ M$_{\odot}$ clouds with $>320$ particles, the artificial dispersal (and formation) of these clouds is likely less important, but worth bearing in mind. 

\subsubsection{Cloud lifetimes}
Because, as we show in the next sections, it is often difficult to establish the precursor or successor of a given cloud, it is very difficult to assign a lifetime to a given cloud. However we can use the continuation of a given cloud to assign its lifetime. We compute a lifetime as the time at which the continuation of a given cloud is at least half the mass of that cloud at $T_0$ (see Sections 5.1.5 and 5.3.1). This is a somewhat arbitrary definition, but does allow us to quantitatively compare clouds. 

\section{Results}
To start, we consider in detail one particular cloud in our simulation, Cloud380. In Section 5.2. we consider several other massive clouds in the simulation, then in Section 5.3 we consider statistics of all clouds over $10^5$ M$_{\odot}$.

\subsection{Example 1: Cloud380}

We consider first one of the more massive clouds found by the clump finding algorithm at time $T_0 = 250$ Myr and name it Cloud 380 (being number 380 on the algorithms list). It is located in a spiral arm 3 kpc from the centre of the galaxy, and has a mass of $2 \times 10^6 M_\odot$, corresponding to 6386 particles. In Figure~3 we show the appearance of the cloud at time $t = T_0 = 250$ Myr (centre panel), and show the evolution of the cloud material over a period $ 230$ Myr $< t < 270$ Myr. The mass of the continuation of Cloud 380 is shown in Section~5.1.5. In the vertical plane, Cloud 380 appears as one coherent structure, centred close to the midplane of the galaxy (see also \citealt{Dobbs2011new} for examples of edge on plots of simulated galaxies).

\begin{figure*}
\centerline{
\includegraphics[scale=0.45, bb=0 0 365 360]{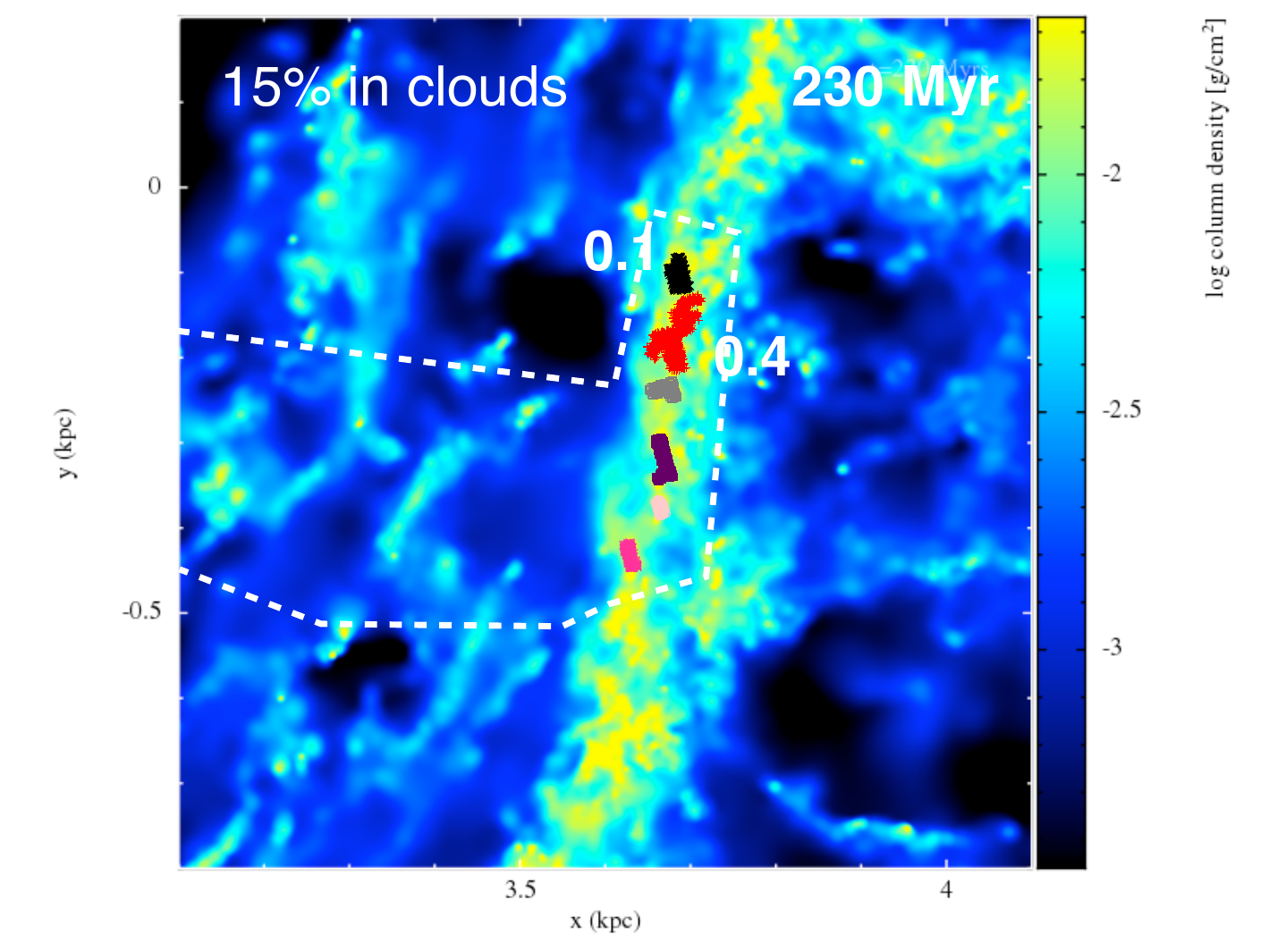}
\includegraphics[scale=0.45, bb=0 0 365 360]{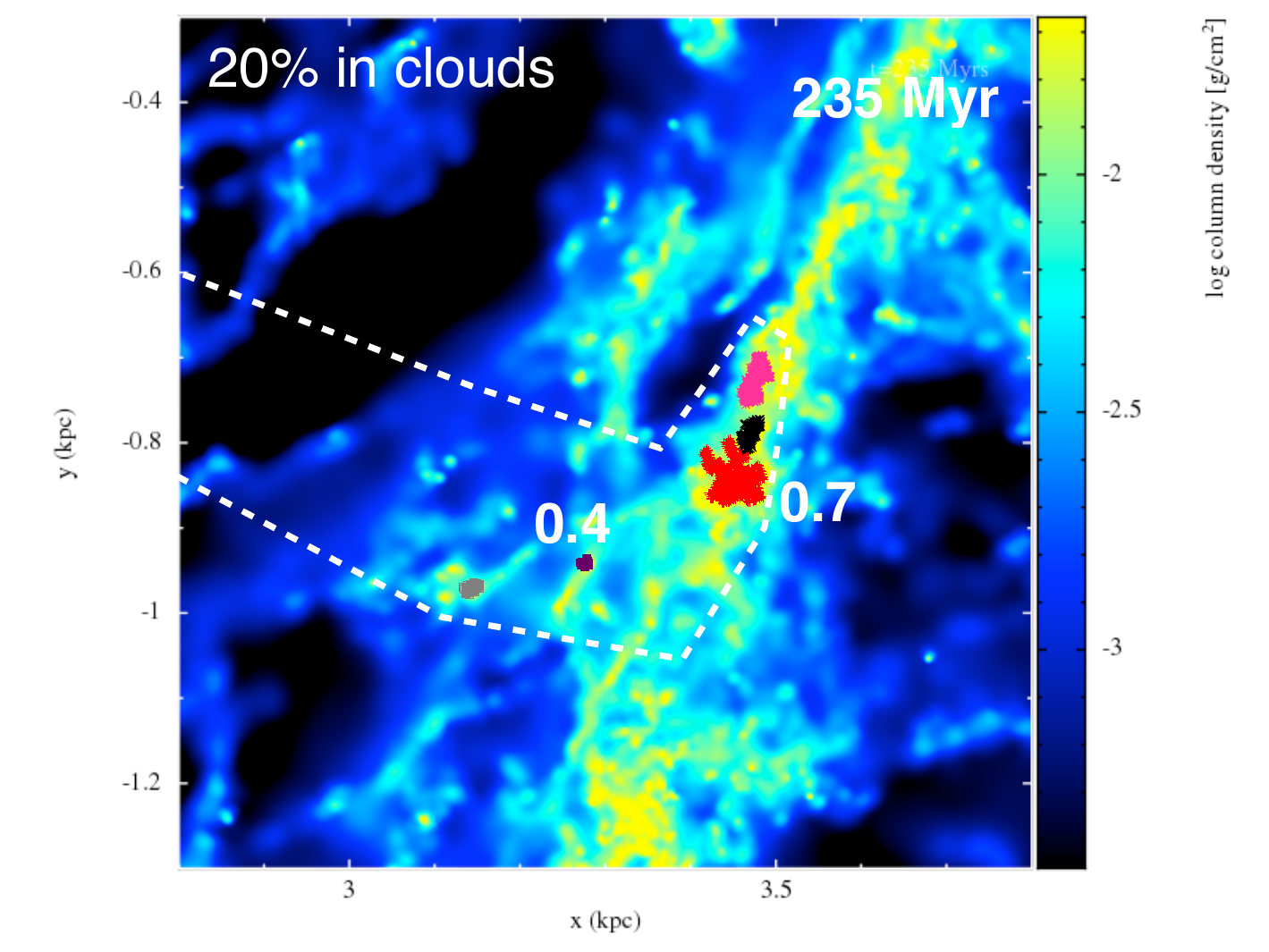}
\includegraphics[scale=0.45, bb=0 0 365 360]{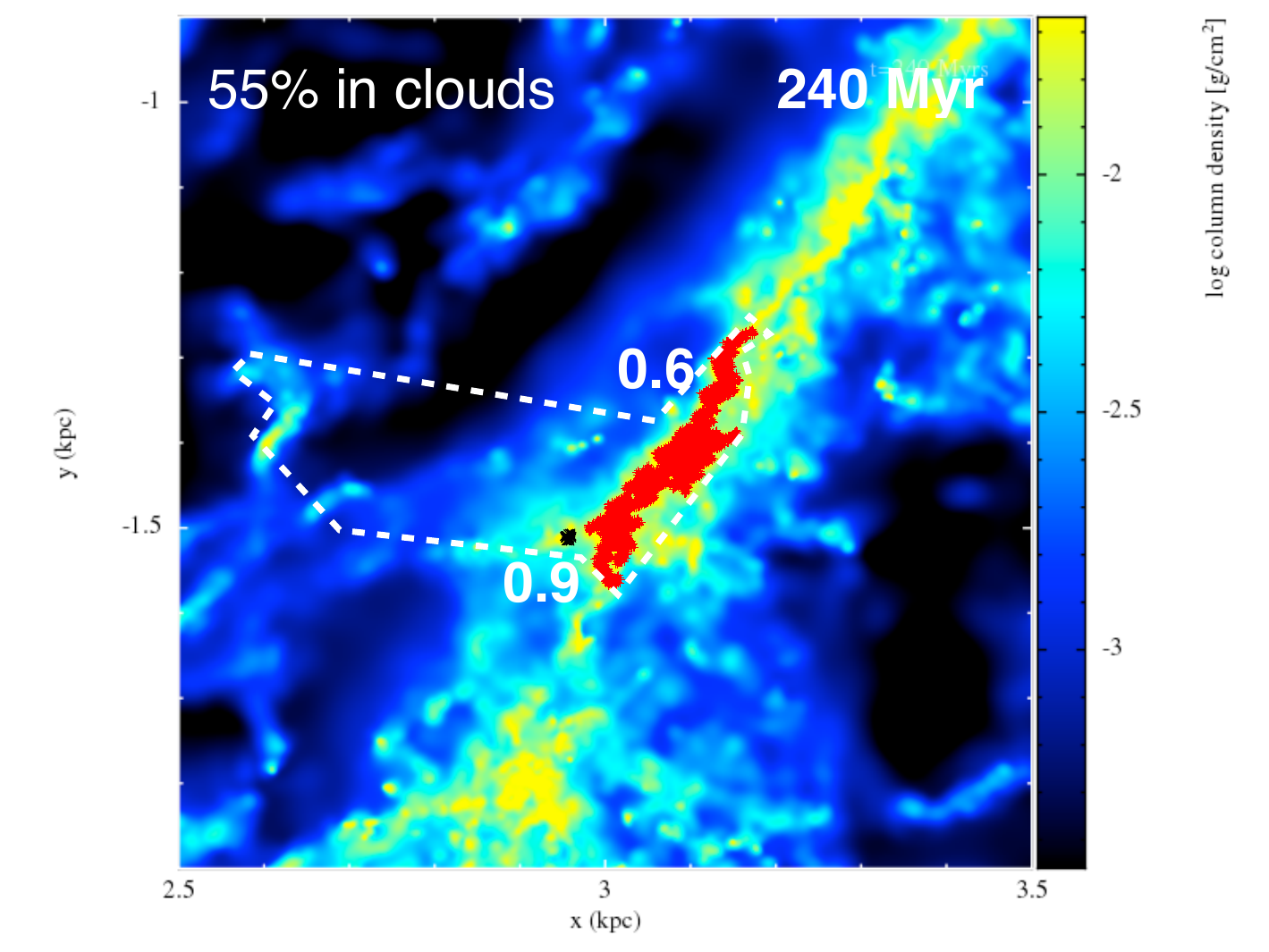}}
\centerline{
\includegraphics[scale=0.45, bb=0 0 365 320]{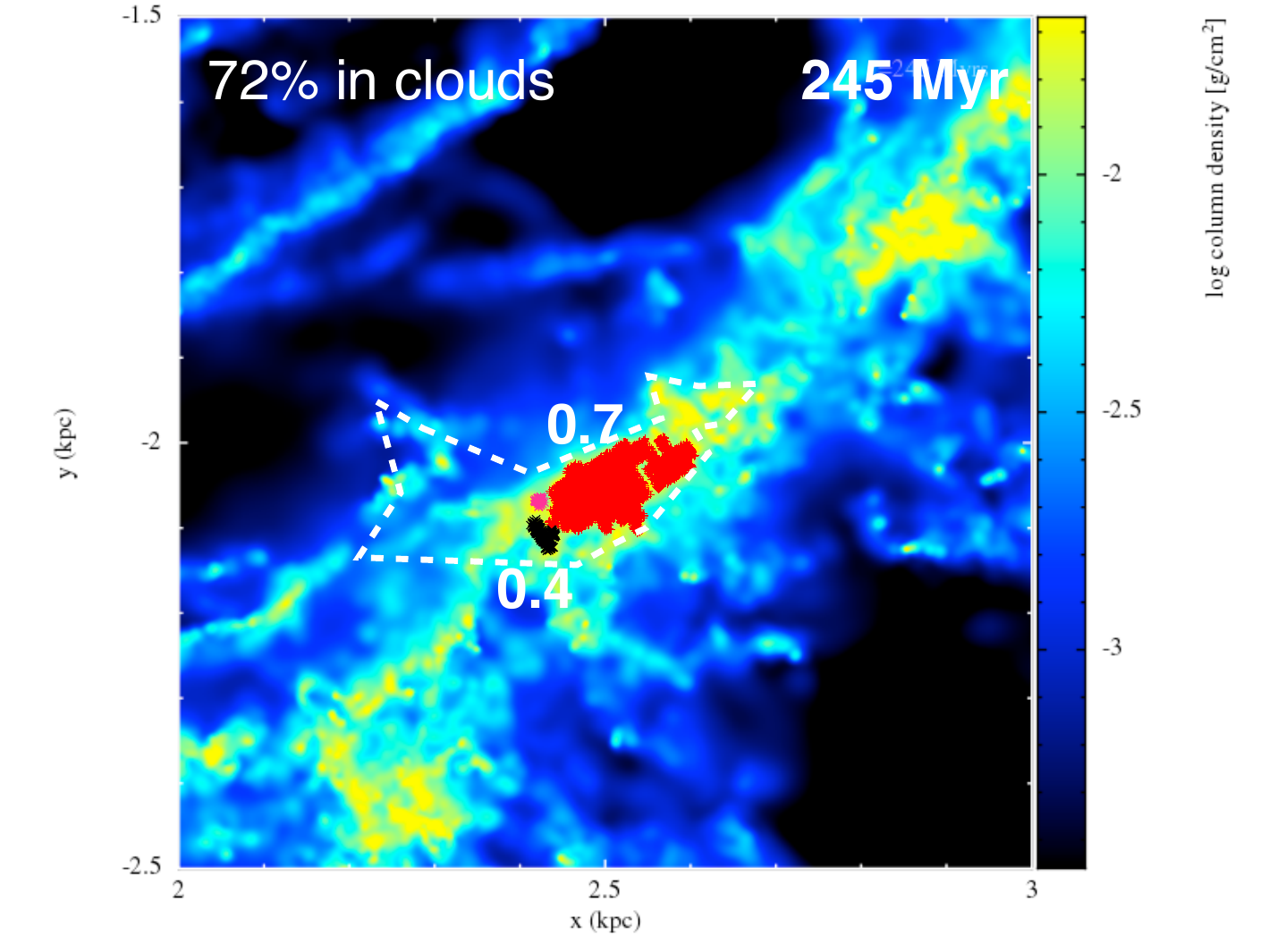}
\includegraphics[scale=0.45, bb=0 0 365 320]{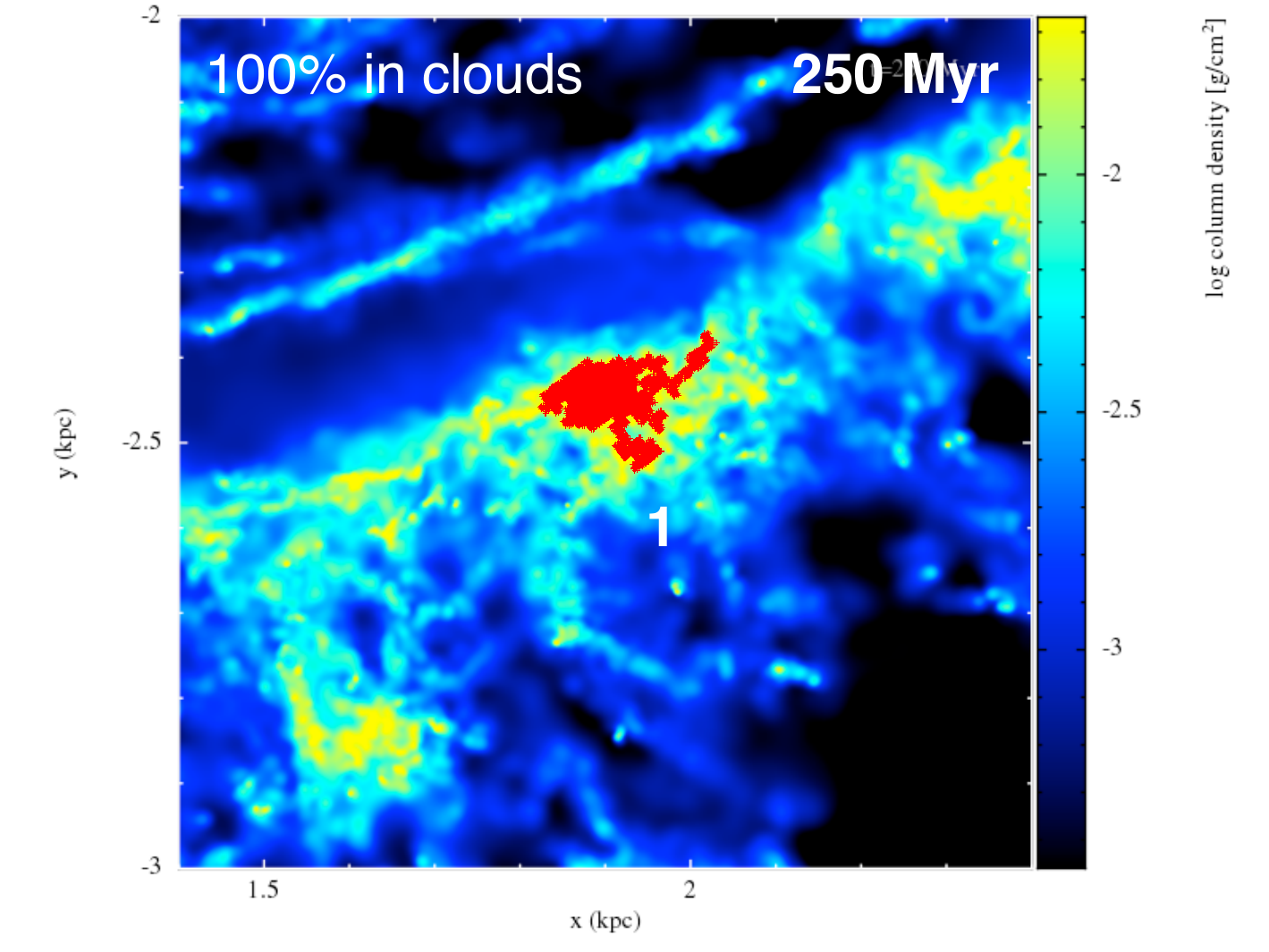}
\includegraphics[scale=0.45, bb=0 0 365 320]{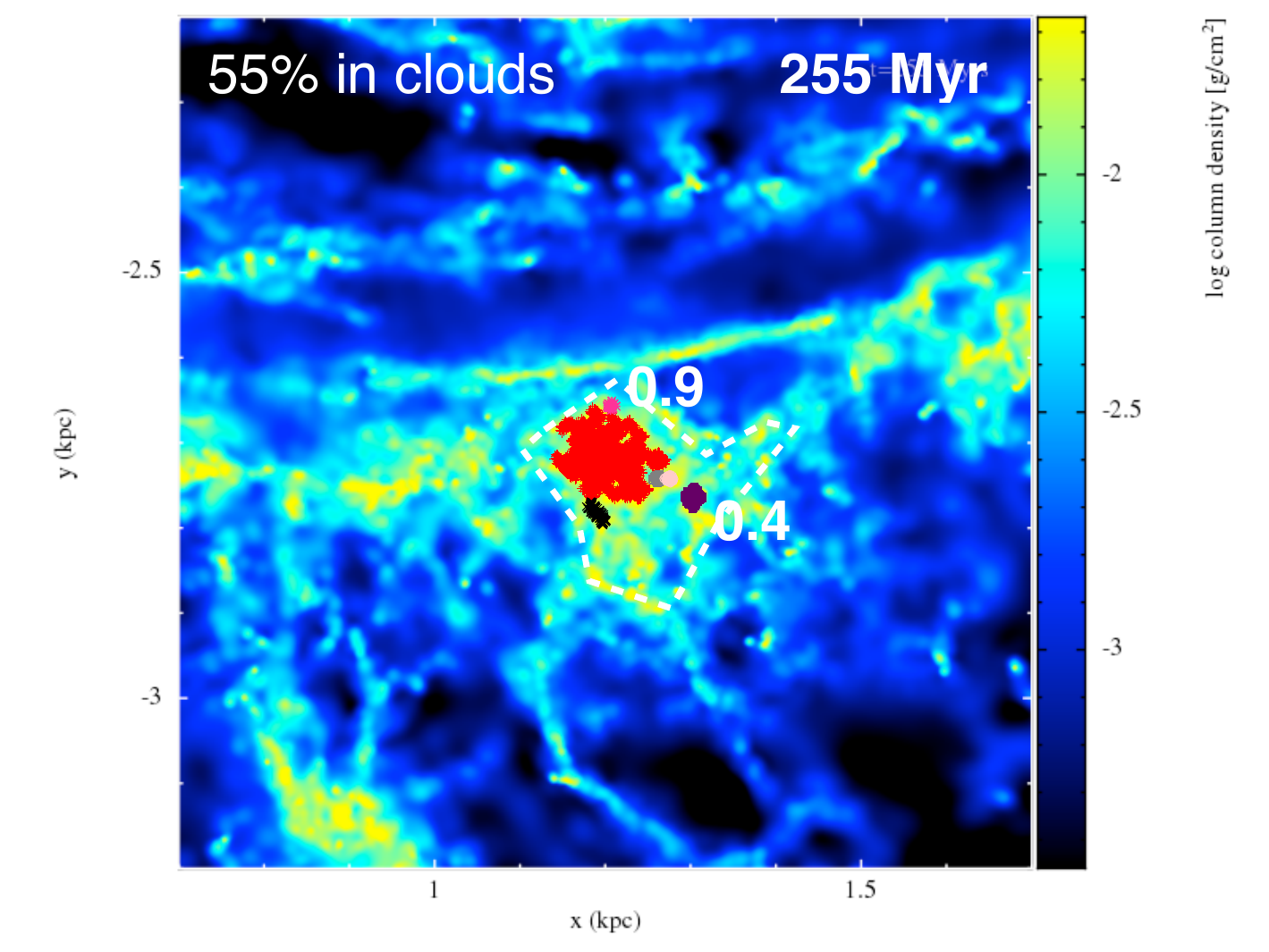}}
\centerline{
\includegraphics[scale=0.45, bb=0 0 365 320]{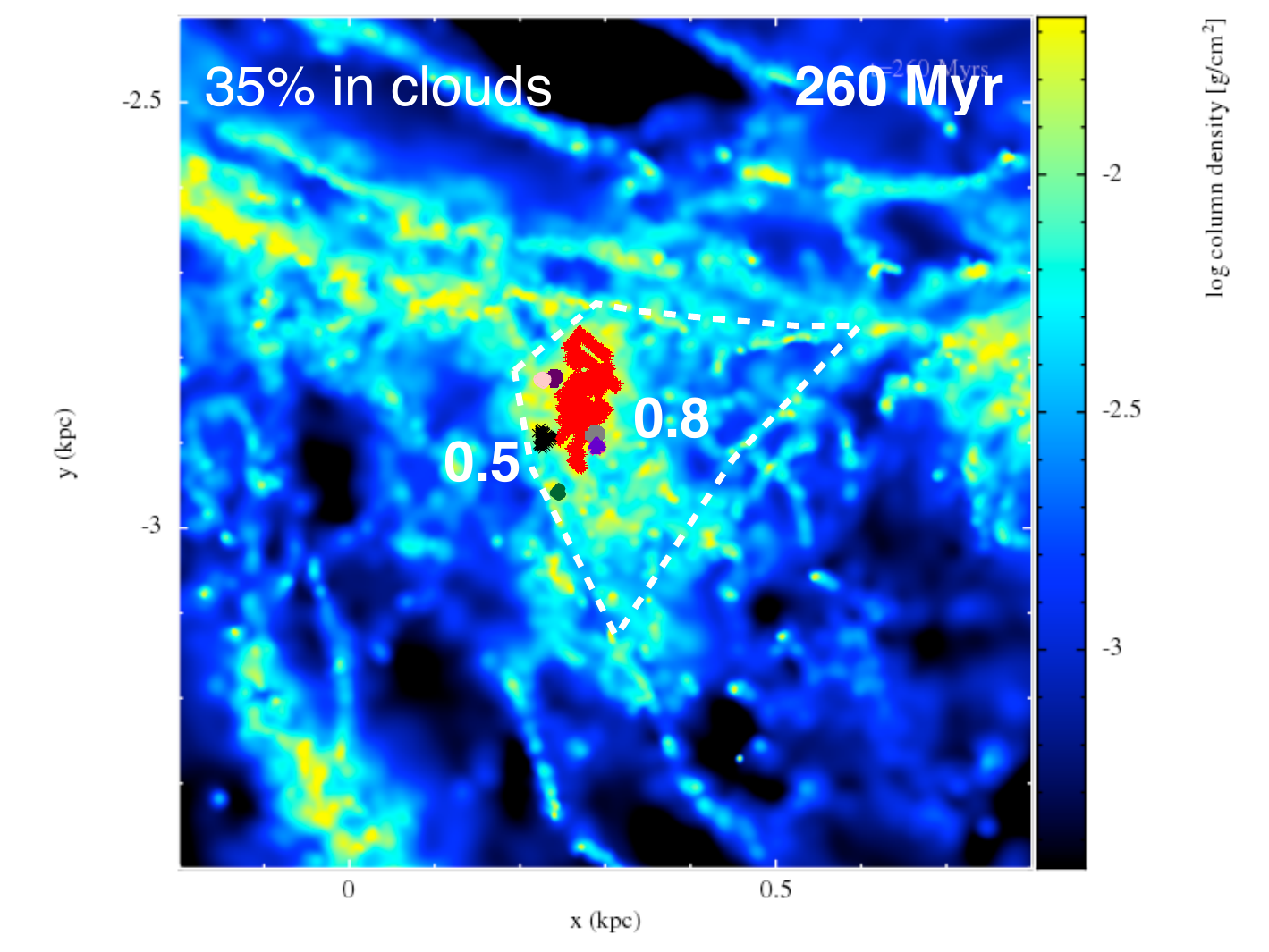}
\includegraphics[scale=0.45, bb=0 0 365 320]{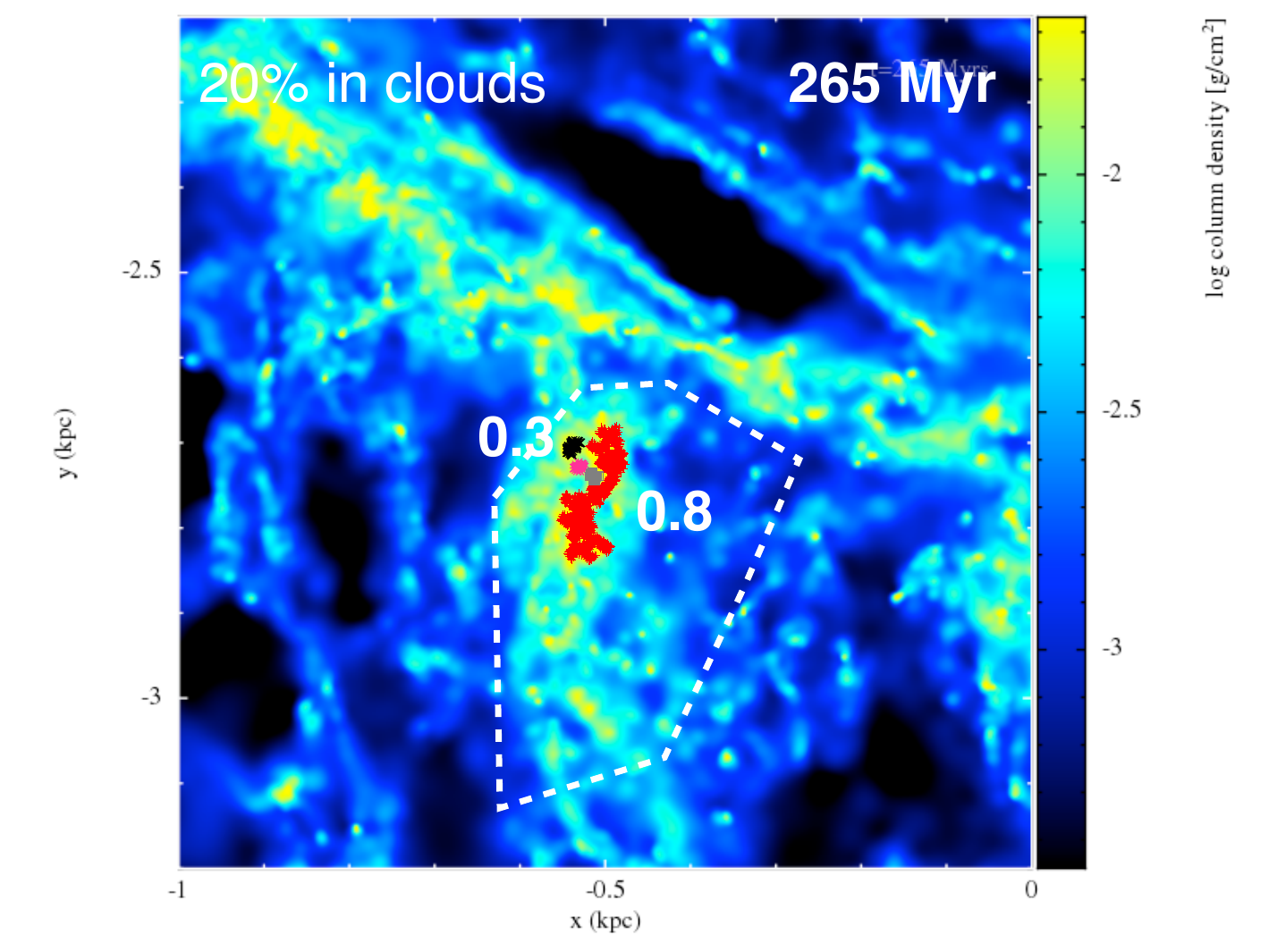}
\includegraphics[scale=0.45, bb=0 0 365 320]{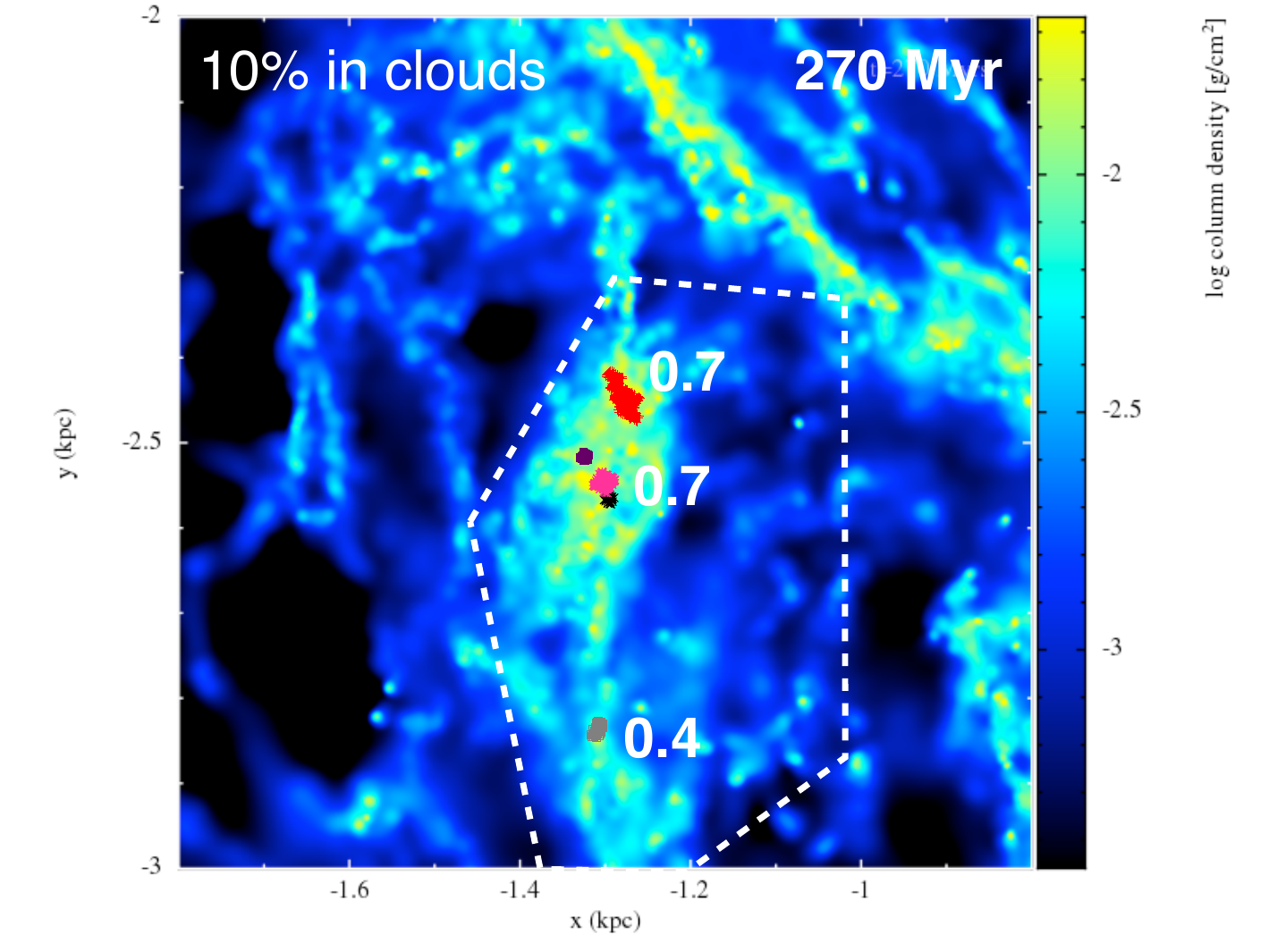}}
\caption{The evolution of Cloud380 is shown over a period of 40 Myr. All the panels are shown in the rotating frame of the spiral potential. Clouds are indicated by blocks of colour overplotted on the column density plots. The clouds shown are all those which contain at least 20 original cloud particles of Cloud380. Cloud380 is shown in the middle panel. The fraction of particles which are original cloud particles of Cloud380 are indicated for some of the clouds, for clarity fractions are typically only shown for a couple of clouds. The total fraction of the original cloud particles of Cloud380 which are located in clouds is also indicated on the top left of each panel. The remainder of this gas lies in the ambient ISM. The locus of the region containing the original cloud particles of Cloud380 is indicated by the dashed white lines for each panel (for the middle panel obviously this locus is simply the region in red).}
\end{figure*}

\subsubsection{Cloud dispersal: evolution at times $t > T_0$}

If we start with the centre panel of Figure~3 and move forward in time, we find that after 20 Myr only ten per cent of the original cloud is still in clouds -- the rest has been returned to the ambient ISM. In each of the later panels we delineate approximately the area covered by the original cloud particles. The major successor cloud is shown in red, but there are also other clouds (shown in different colours) which contain a substantial number ($> 20$) of the original cloud particles. The numbers by some of these clouds shows what fraction of each of those clouds is made up of original cloud particles. Thus we see that after $T_0$ Cloud380 fragments into multiple smaller clouds, although there is still one massive cloud evident. The cloud disperses through a combination of shear and feedback (We discuss the role, and relative importance, of these processes below.).

\subsubsection{Clouds, spurs and the effects of stellar feedback}
Until $T_0=$ 250 Myr, Cloud380 (or rather the progenitor of Cloud380) experiences net growth as it travels along the spiral arm. However after 250 Myr, there is not much gas in the spiral arm in the vicinity of Cloud380, and Cloud380 then leaves the spiral arm, hence its mass decreases. Not long after $T_0$, for 255 Myr $< t < $ 270 Myr, the whole of the successor of Cloud380 lies in a `spur' feature downstream of the arm. At times from 255 to 265 Myr, the successor of Cloud380 is the largest cloud within a 1 kpc radius by a factor of a few. In particular there are no comparable mass clouds in the spiral arm. After 270 Myr, material begins to build up in the spiral arm again. At 270 Myr, the most massive GMCs in the arm and spur are comparable ($\sim 2 \times 10^5$ M$_{\odot}$). 

\begin{figure*}
\centerline{
\includegraphics[scale=0.45, bb=0 0 365 360]{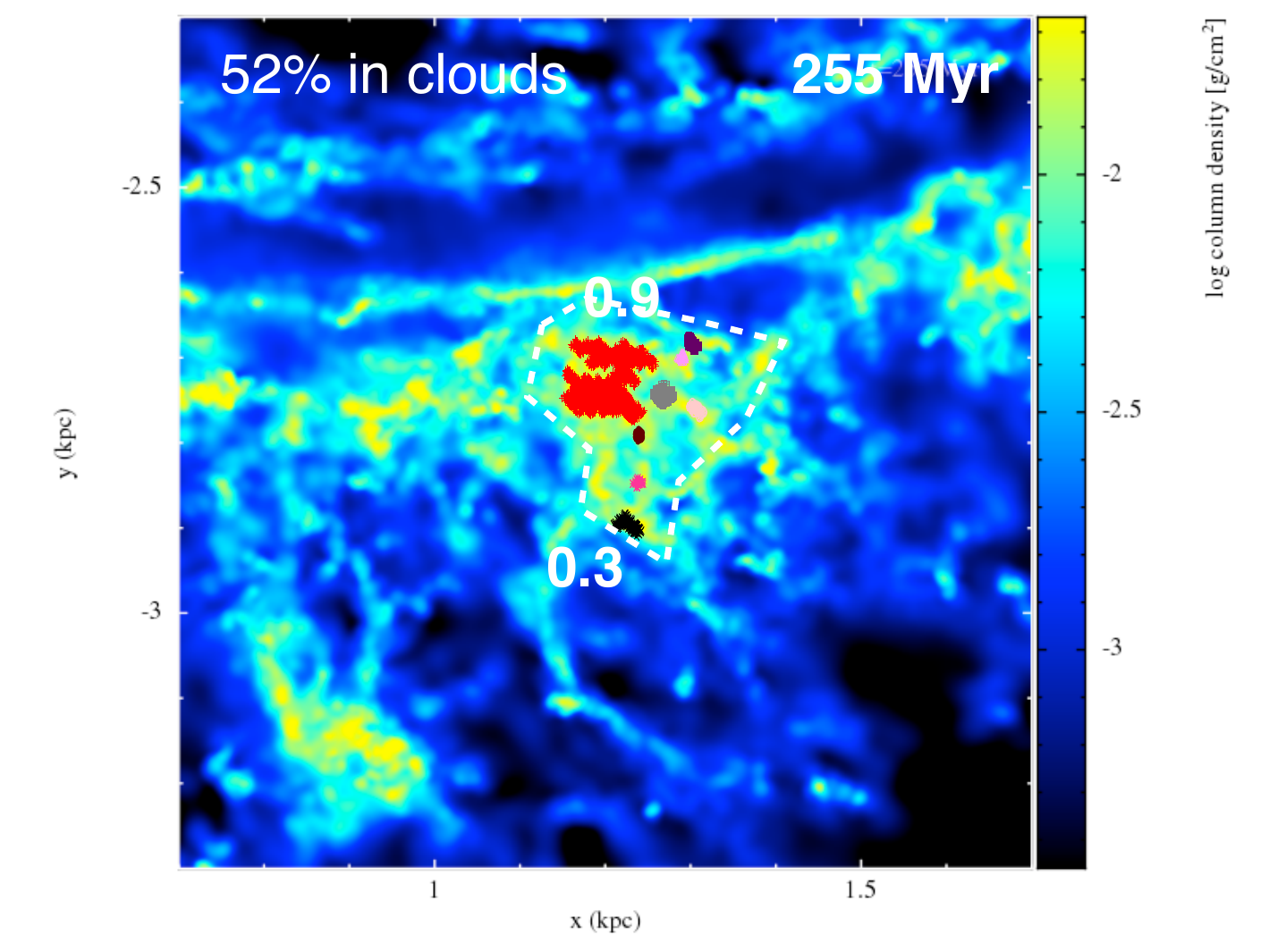}
\includegraphics[scale=0.45, bb=0 0 365 360]{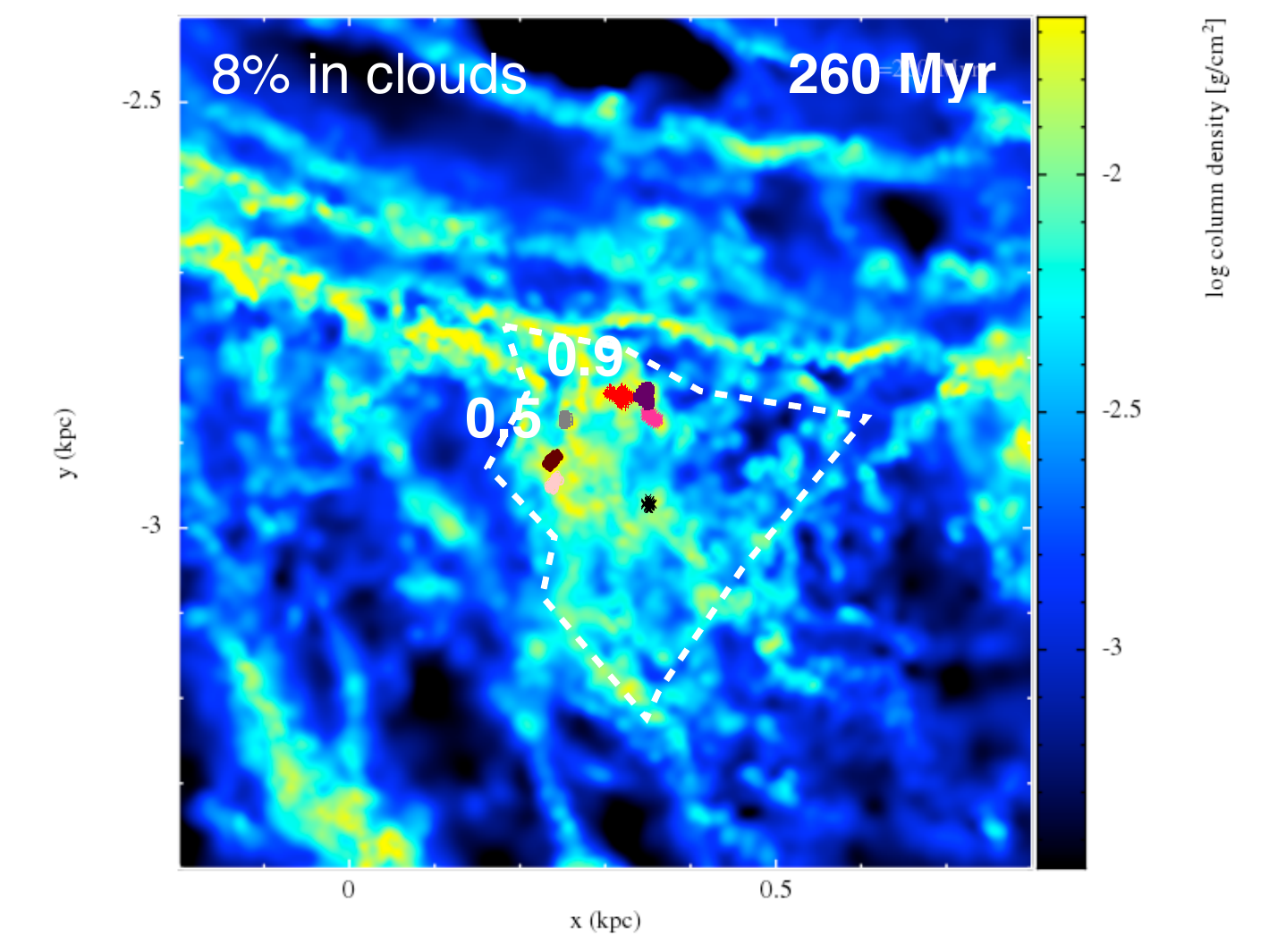}
\includegraphics[scale=0.45, bb=0 0 365 360]{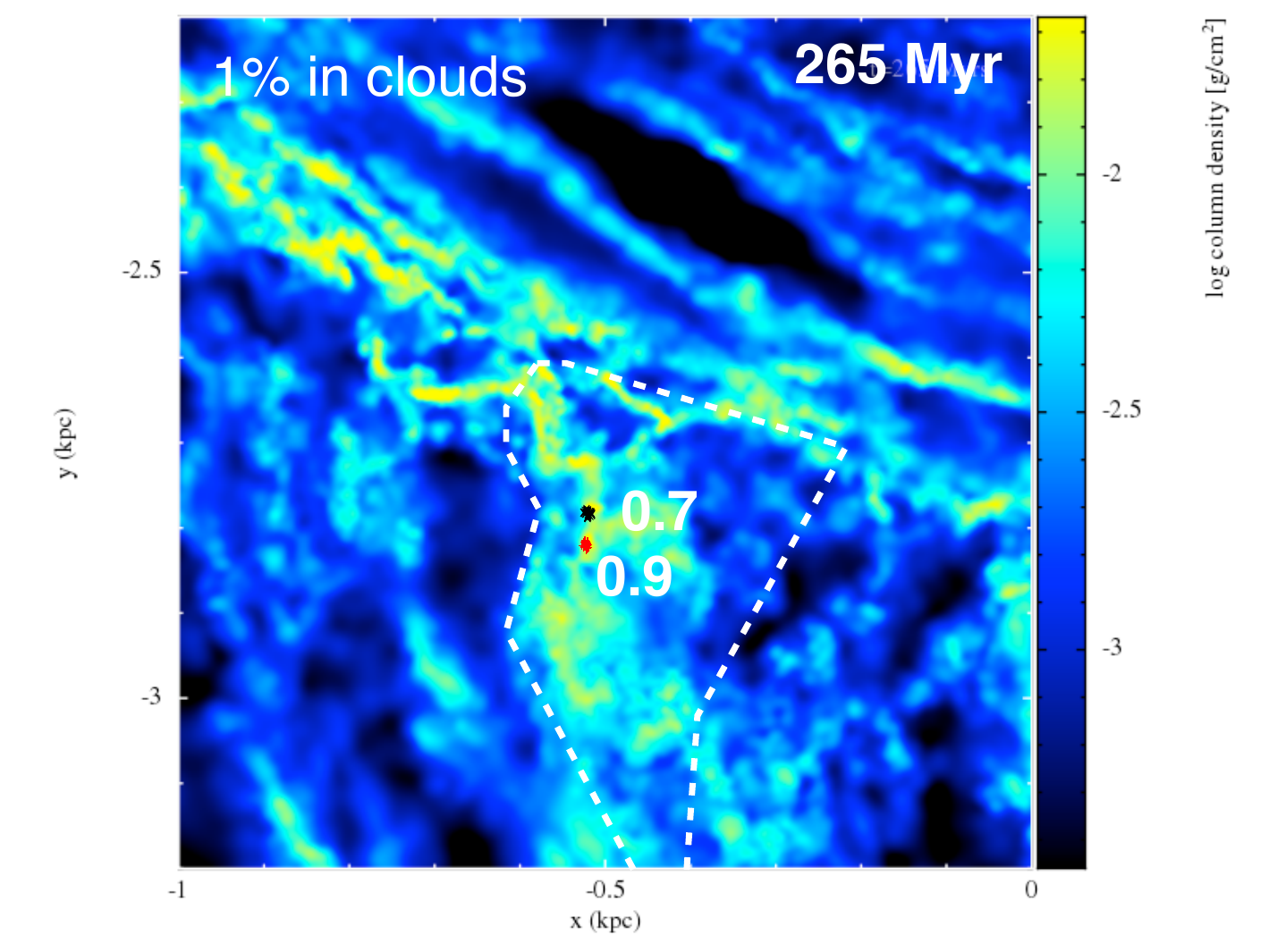}}
\caption{The corresponding panels are shown for the times of 255, 260 and 265 Myr shown in Figure~3, but for a simulation ran from $T_0=$250 Myr without feedback (or self gravity). The large scale structure of the gas is very similar without feedback, suggesting it is shear rather than feedback which drives cloud disruption. Feedback appears more relevant for disrupting clouds on scales smaller than GMCs.}
\end{figure*}

This picture can be compared with the results of \citet{Schinnerer2013} where they study clouds and spurs in a region of M51. They find that the most massive clouds are preferentially located on the downstream side of the arm. For their particular section of a spiral arm,  \citet{Schinnerer2013} also find no substantial, massive star-forming clouds within the spiral arms. We note that this is also seen in the simulations, but only at particular times and locations. For example at times 260 Myr and 265 Myr (Figure 3), we see in our chosen arm section massive clouds situated in spurs, whilst the gas has not yet had time to re-form massive clumps in the spiral arm (This again reflects our finding the molecular clouds in spiral arms are not `ridge clouds').

Stellar feedback contributes to the structure of the gas along the spiral arms, and in the inter-arm regions, although clear holes relating to multiple supernovae are not easy to see. This is partly because we choose a moderate star formation efficiency (for example in \citealt{Dobbs2011new} with a star formation efficiency of 20\% we could clearly see large holes). There is substructure due to feedback within, and around, the clouds in our simulations but it is on relatively small scales (e.g. 10-20 pc) and would require zooming in to the plots in Figure~3 to see clearly. \footnote{Note that we have chosen to plot the cloud components as opaque blocks of colour in order to emphasise particular subsets of cloud particles, but that this then tends to obscure the detailed structure of the cloud itself.}

\subsubsection{Cloud formation: evolution at times $t < T_0$}

Prior to time $T_0 - 20$ Myr less than 15 percent of Cloud380 is in cloud gas, with the rest in the ambient ISM. In Figure 3 we plot at early times (and also late times) the area containing most of the particles which find their way into Cloud380. We can see that at time $T_0 - 20$ Myr, most of the gas which eventually forms Cloud380 lies in the interarm region, and that the interarm gas is generally diffuse, rather than lying in clouds. There are a few interarm clouds which enter the spiral arm, as seen at 235 Myr, but generally the interarm gas is not dense enough or not sufficiently concentrated (recall we have a minimum cloud mass of  around $10^4$ $M_\odot$) to become apparent as clouds.

\subsubsection{Does shear or stellar feedback cause Cloud380 to disperse?}

We see that after $T_0=250$ Myr, Cloud380 splits up into a number of smaller clouds, and by 270 Myr, a successor of Cloud380 is not clearly distinguishable. 
Since the virial parameter of Cloud380 is $\alpha=3$, in this instance the cloud is unbound and will naturally expand without feedback.
However bound regions within Cloud380, and other bound $>10^5$ M$_{\odot}$ clouds in the simulation (see Figure~A1 in the Appendix and Table~1) cannot be dispersed simply by the dynamics of the clouds at $T_0$. Also, internal motions which keep clouds unbound, likely originate at least in part from feedback and travel through the spiral arms \citep{Dobbs2011new}. 
Overall it is likely that Cloud380 splits up overall due to some combination of shear and feedback. We argue here that shear is responsible for the large scale dispersal of the cloud, but that feedback is also likely to break up the densest parts of Cloud380.

One indication that shear is likely to play a role in the dispersal of Cloud380 is simply the morphology of the region at later times, and how gas is stretched into a spur. In Figure~4 we show panels from a resimulation from 250 Myr where we did not include stellar feedback (although in order to do this we also were not able to include self gravity).  Figure~4 shows panels from the resimulation, for the three times of 255, 260 and 265 Myr (equivalent to those in Figure~3). The evident similarity of the morphology of the region with and without feedback suggests that at least on larger scales, it is shear which disrupts the massive clouds and causes them to break up into smaller clouds (although because self gravity is not included the clouds have dispersed more in Figure~4 compared to Figure~3). 

We also consider quantitatively the likely role of shear for Cloud380. In Figure~5, we compute the surface density at which the tidal force from the galactic potential becomes comparable to the self gravity of the gas. More specifically, we compute the critical surface density, $\Sigma_{crit}$ such that
\begin{equation}
r_{cloud} \frac{dF}{dr}=\Sigma_{crit} G
\end{equation}
where $dF/dr$ is the gradient of the force across the cloud due to the gravitational potential, per unit mass. For Cloud380 we took $r_{cloud}=85$ pc. Figure~5 indicates that at the galactic radius of 3 kpc, the surface density at which shear becomes important ($\sim90$~M$_{\odot}$~pc$^{-2}$) is comparable to the surface density criterion for the clouds ($100$~M$_{\odot}$~pc$^{-2}$). We also show on Figure~5 the timescale associated with the shear at different galactic radii. We estimate this timescale as the inverse of the Oort's constant $A$, where
\begin{equation}
A=\frac{1}{2}\bigg(\frac{V}{R}-\frac{dV}{dR}\bigg).
\end{equation}
At $R=3$ kpc, this timescale is $\sim30$ Myr. As we see in Figure~3, this is not too dissimilar for the time for the region to be sheared into a spur.
Previous authors have attempted to use a Toomre, or Toomre-like criterion to determine the relevance of shear in a molecular cloud \citep{Hunter1998,Dib2012}. However this is only applicable for an axisymmetric disc, and there is no equivalent local criterion. \citet{Dib2012} conclude that shear is not important for clouds in the molecular ring, however they used a different criteria, and obtained slightly lower values of $A$ than found in our model.

Finally we also examined the role of shear and feedback by examining the distribution of surface densities within Cloud380 and its successors (Figure~5, lower). The most dense regions disappear very quickly after 250 Myr, likely due to feedback processes. However on longer timescales, of few tens of Myr, the whole distribution of $\Sigma$ is shifted to low surface densities. Again this likely corresponds to the whole cloud being sheared out to lower surface densities. 

We note that whilst at $R\sim3$ kpc, shear appears to be important at disrupting GMCs, at larger radii, this may not be the case. At larger radii, $\Sigma_{crit}$ is much lower, indicating that dense clouds are unlikely to be disrupted by shear. Likewise, the timescale for disruptions increases to several tens of Myr. We consider other clouds in Section 5.2

\begin{figure}
\centering
\includegraphics[scale=0.35]{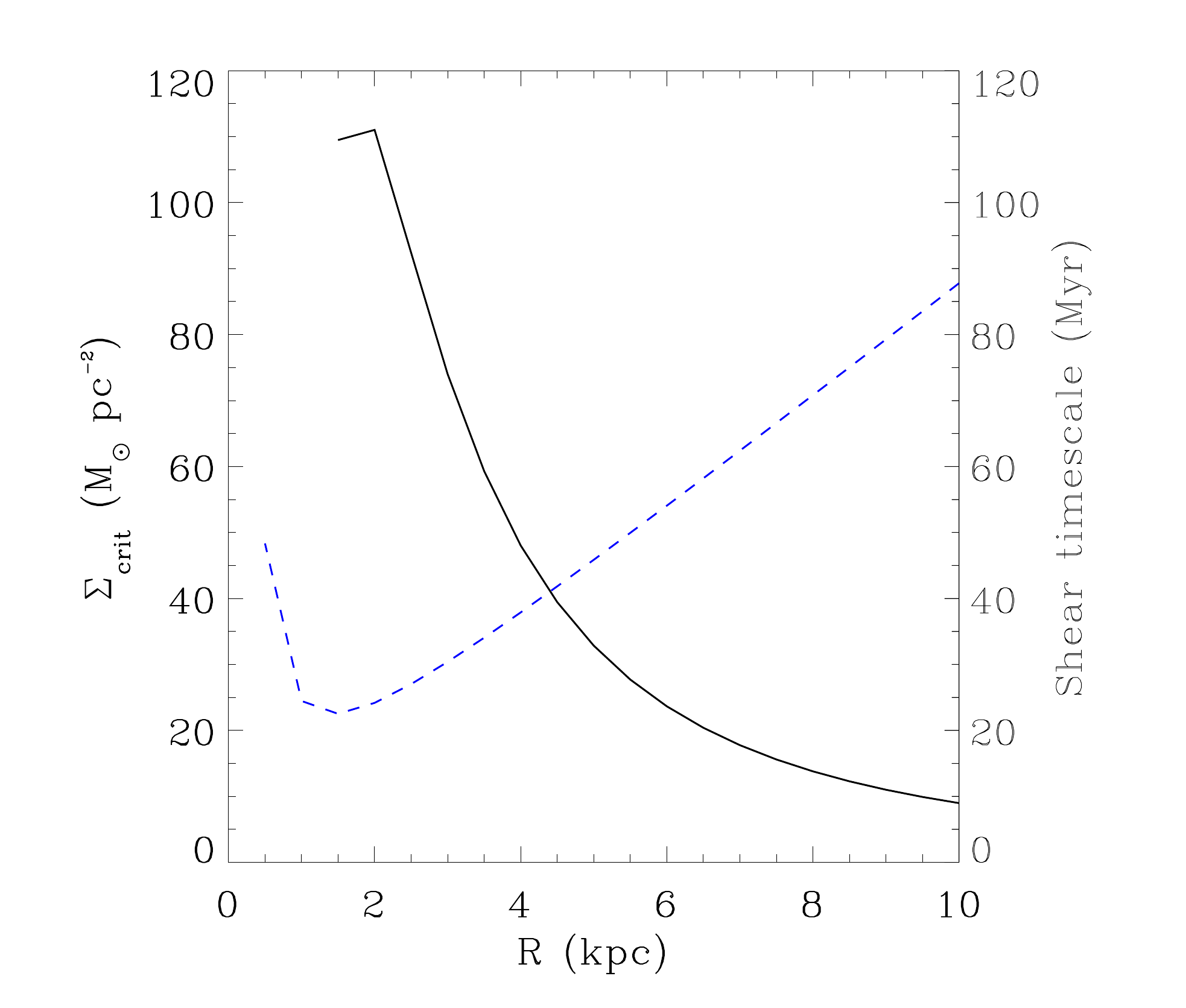}
\includegraphics[scale=0.37]{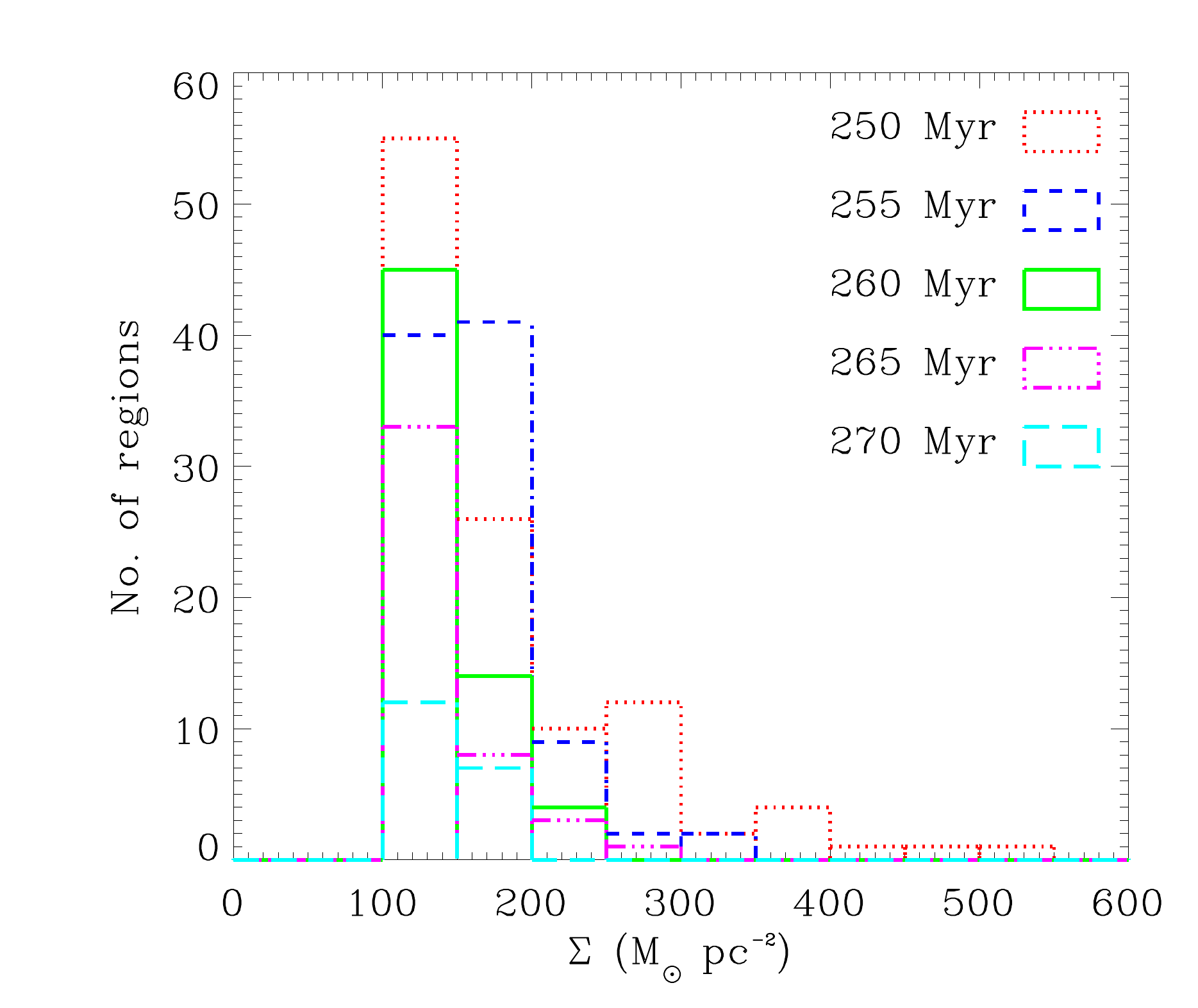}
\caption{The top panel shows the critical surface density for shear to be effective (calculated according to Eqn~2), plotted against galactic radius (solid black line). The shear timescale, calculated as the inverse of Oort's constant $A$ is plotted as the blue dashed line. Cloud380 is situated at a radius of $\sim$ 3 kpc. The lower panel shows the distribution of surface densities in 10pc $\times$ 10 pc boxes over all clouds which contain 20 or more original particles of Cloud380, from times of 250 Myr (i.e. for Cloud380) to 270 Myr. There is an immediate disappearance of the highest density regions from 250 to 255 Myr, likely due to feedback. There is then a more gradual decrease across all surface densities, likely due to shear. }
\end{figure} 

\begin{figure}
\centering
\includegraphics[scale=0.43]{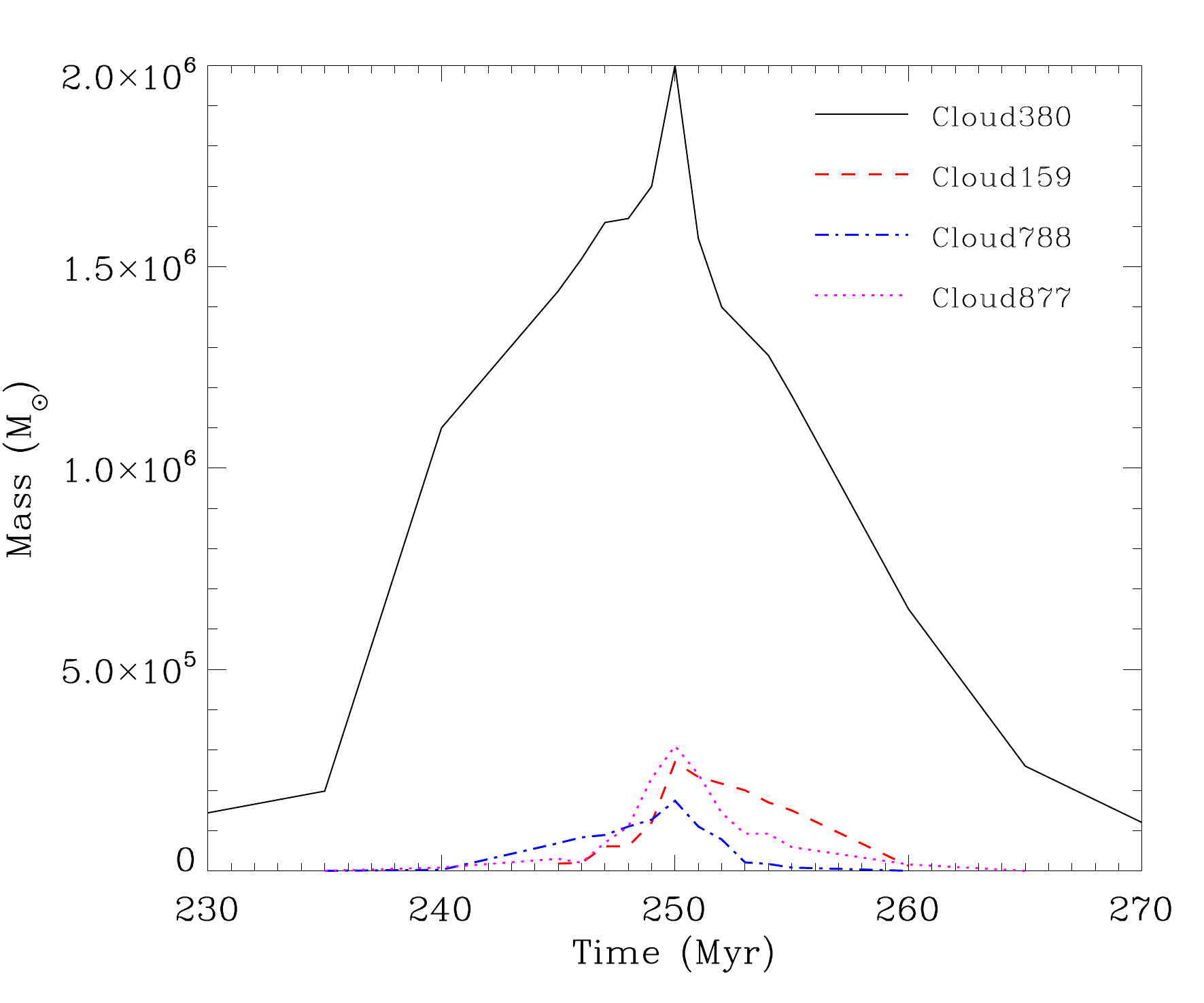}
\caption{The mass of the continuation of Cloud380 (the largest mass of original cloud particles situated in a precursor or successor of Cloud380), and several smaller clouds is shown versus time. The time resolution is 1~Myr between 245 and 255~Myr, and 5~Myr otherwise.}
\end{figure} 

\begin{figure}
\centering
\includegraphics[scale=0.43]{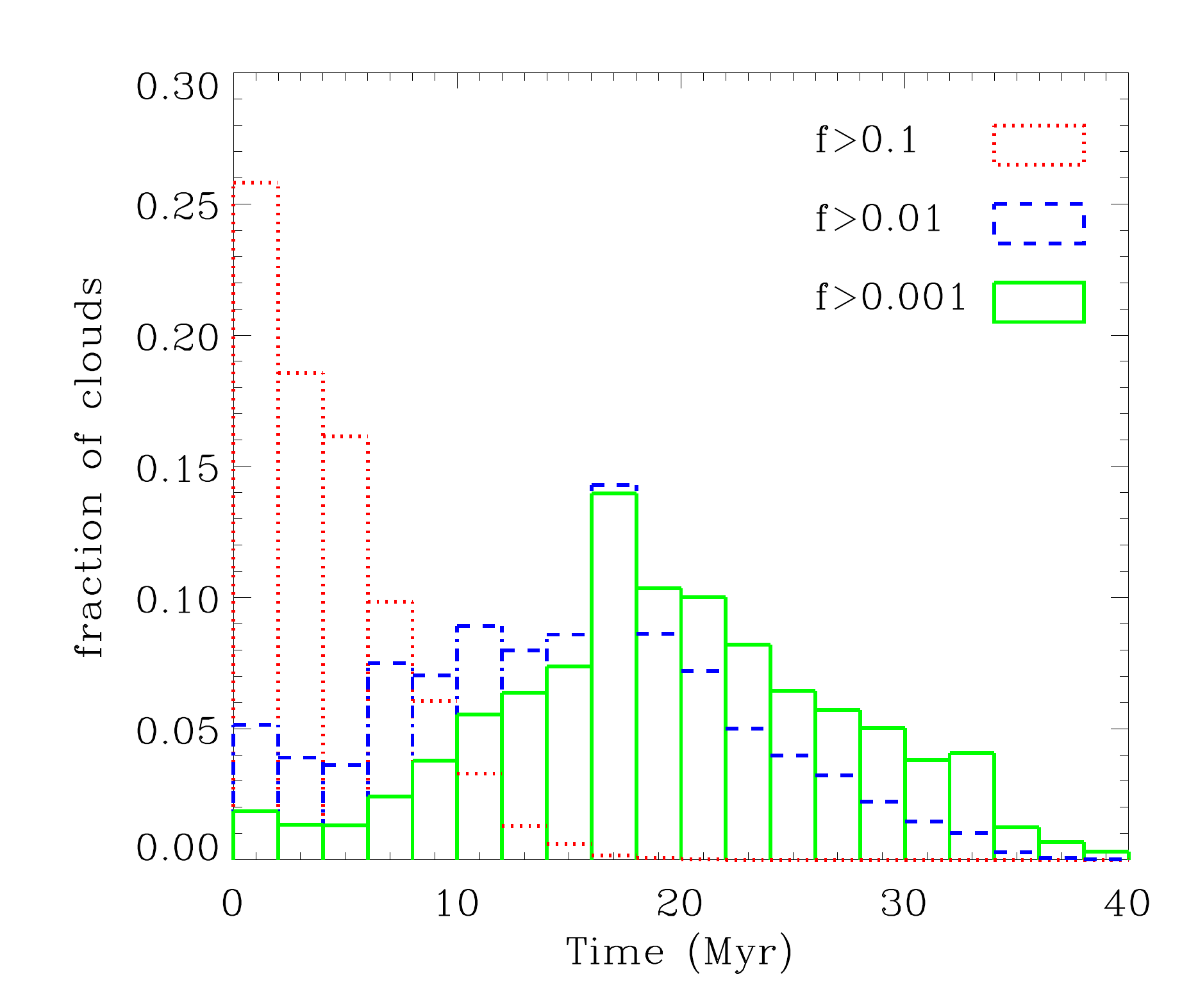}
\caption{This figure shows the distribution of times the original gas particles of Cloud380 remain above a given molecular gas fraction.}
\end{figure} 

\subsubsection{Cloud lifetime}
Figure~3 demonstrates that assigning a lifetime to Cloud380 is difficult. Cloud380 comprises a number of smaller clouds at earlier and later times, and furthermore these clouds do not contain all the same particles as Cloud380. So for example at 235 or 270 Myr there is no obvious counterpart to Cloud380. We first estimate a dynamical lifetime based on the time for the cloud mass to change. Figure~6 shows the change in mass of the continuation of Cloud380. Recall that the continuation of a cloud is the cloud which contains the greatest number of original cloud particles. In Figure~6 we show the mass of the original cloud particles in these clouds.
Taking our definition of the lifetime as the time over which there is an object which has at least half of the \textit{same} mass of Cloud380 (Section 4.3.3), then the lifetime of Cloud380 is $\sim$20 Myr.

An alternative lifetime could be based on the time gas remains molecular, i.e. based on the chemistry of the cloud rather than the flow of gas. In Figure~7 we show the time after 250 Myr for which gas exhibits a minimum molecular fraction, for the original cloud particles of Cloud380.
Because, due to resolution, we insert feedback at comparatively low densities, the most dense gas in our calculation is only 10's or 100's cm$^{-3}$. Gas at the former densities is typically not fully molecular (see Dobbs et al. 2008). Gas retains a high molecular fraction for only a short time - reflecting that  gas does not remain very long at densities of $>$100 cm$^{-3}$ (Dobbs et al. 2012), likely due to feedback. Gas retains high molecular fractions for up to $\sim$ 10 Myr.  Gas which exhibits molecular fractions $>0.01$, thus $>$10 cm$^{-3}$, retains this fraction for typically $\sim$ 16 Myr, which is more in agreement with the timescales found from dynamical arguments. The timescale associated with the molecular gas is very likely to depend on the surface density of gas in the galaxy and the amount of molecular gas. For example, for M51, which is predominantly molecular in the centre \citep{Rickard1981,Scoville1983} a GMC lifetime based on how long gas remains molecular will likely be long, whilst a dynamical lifetime would likely be shorter (i.e. arm GMAs/GMCs are not the same clouds as inter-arm GMCs seen at some time later).

\subsubsection{Star formation in Cloud380}

In Figures~8 and 9 we show the star formation which takes in place in \textit{the gas particles} which constitute Cloud380, i.e. the original cloud particles of Cloud380. Although we do not include star particles in the simulations, we do record when, where and how many stars form each time we insert feedback. Figure~8 (upper panel) indicates that star formation is much higher for the gas between $\sim$240 and $\sim$260 Myr, in agreement with our very approximate maximum cloud lifetime of 20 Myr. The total mass of stars formed is $\sim 6 \times 10^4$ M$_{\odot}$, indicating that the star formation efficiency of Cloud380, measured as the total mass of stars formed (over $\geq$ 20 Myr) divided by total mass of the gas is around 2.5 \%. 

The instantaneous star formation efficiency, although not a true efficiency, is an observable measure of the star formation activity in a cloud, taken as the mass of young stars present divided by the mass of gas. The instantaneous efficiency in nearby clouds (Lupus, Taurus, Orion, Ophiuchus) is observed to range from $\lesssim0.3$~\% to 10 \% \citep{Huff2006,Merin2008,Mizuno1995,Hatchell2012}. By comparison, at 250 Myr in our simulation, the instantaneous efficiency is $\sim1$ \%.

In Figure~9 (top) we show the age distribution of stars formed from the original cloud particles of Cloud380 at a time of 250 Myr to a maximum age of 20 Myr. Without star particles, we are not able to tell when stars decouple from the gas. Instead our results represent the maximum ages of stars present in the clouds. Our results agree with \citet{Hartmann2012} who find that clouds typically have many more very young stars. Our distribution extends to higher ages however, as we take a much more massive cloud. Tracing the star particles will only make this distribution steeper. Changing the feedback scheme (e.g. including a delay) could potentially allow more older stars to be present. However the timescale for massive star formation is likely to be short, and whilst supernovae  occur at the end of a massive star's lifetime, stellar winds are present throughout.

\begin{figure}
\includegraphics[scale=0.34]{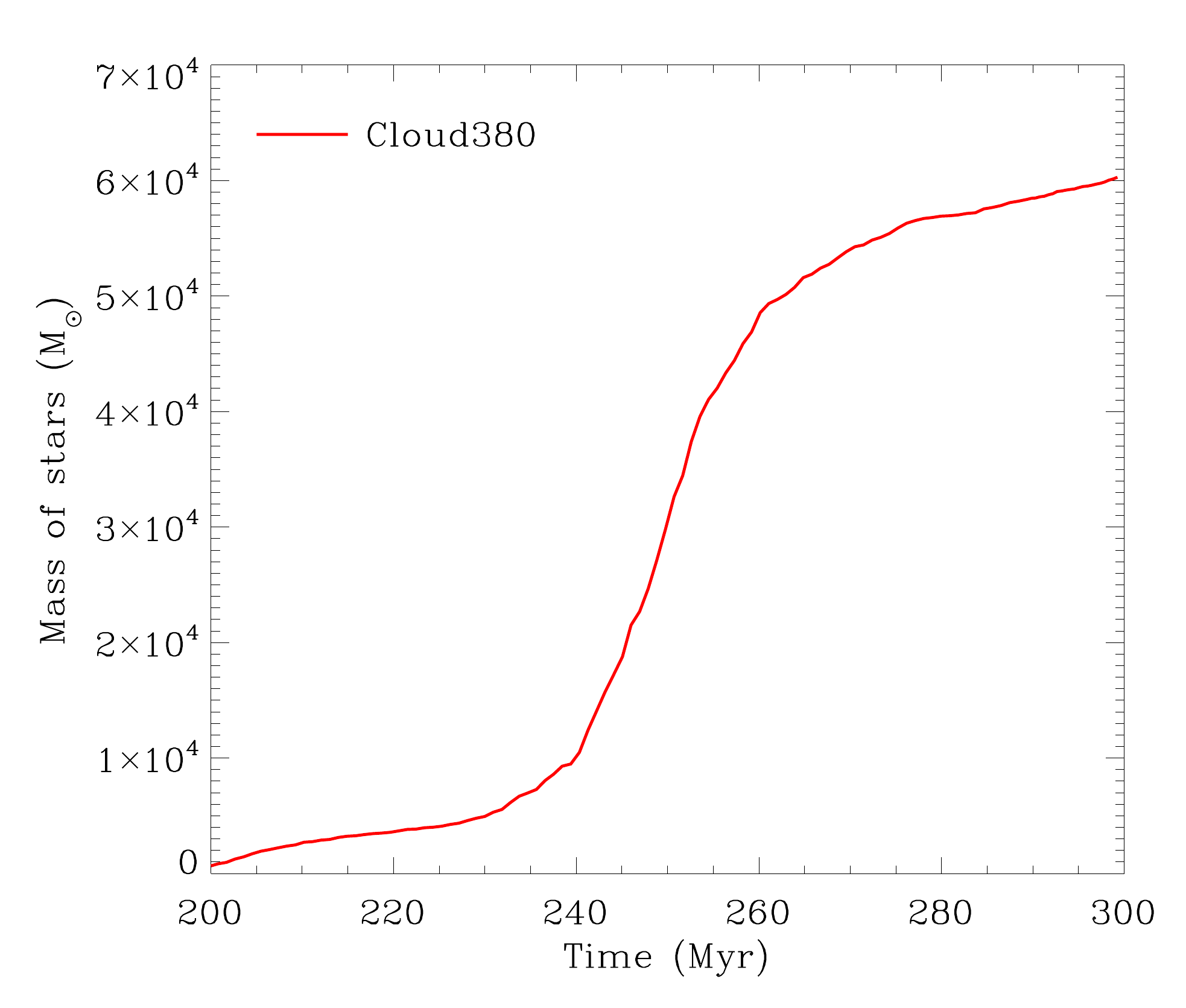}
\includegraphics[scale=0.34]{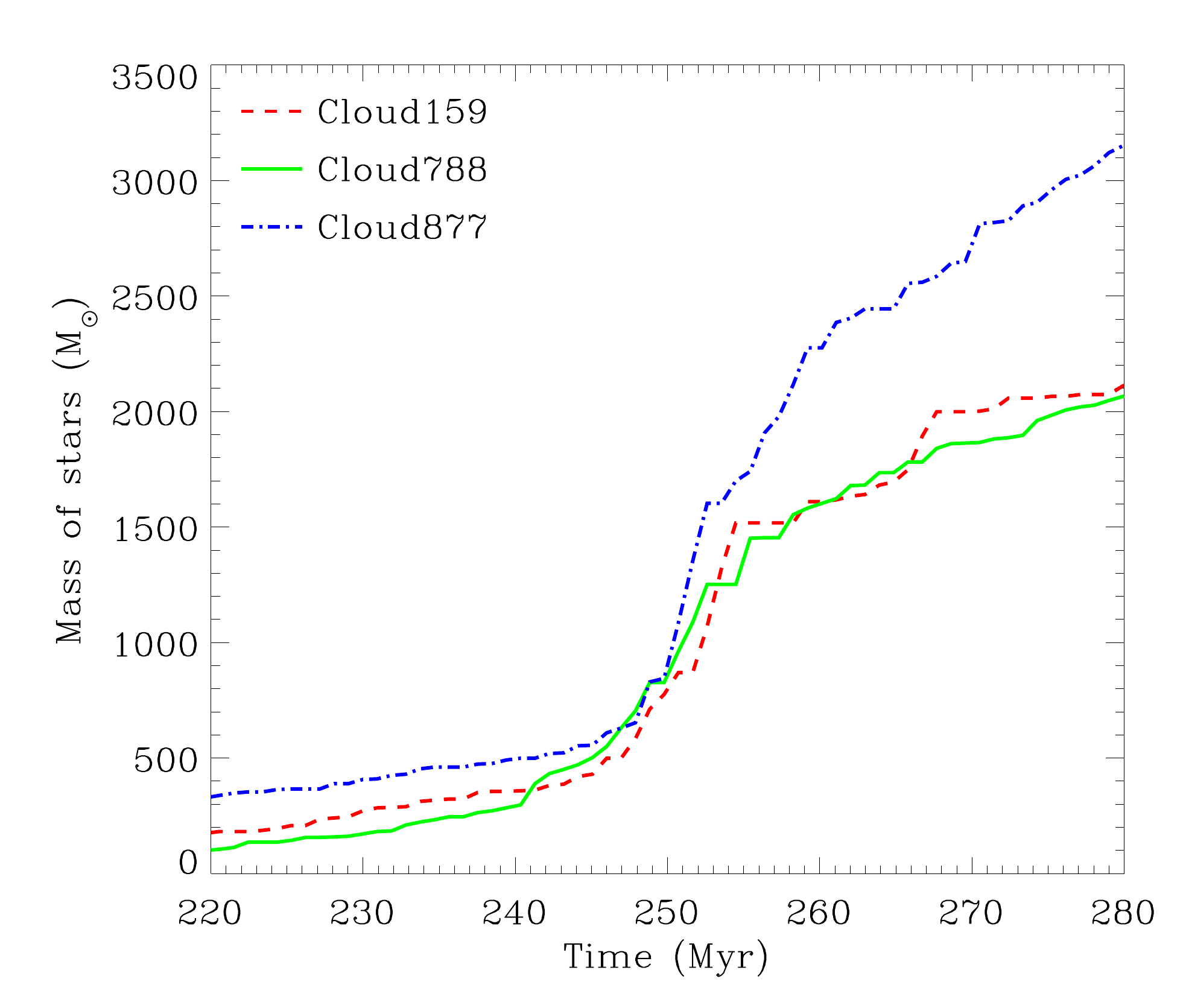}
\caption{The top panel shows the cumulative mass of stars formed versus time for Cloud380 (top), which has a mass of $2 \times 10^6$ M$_{\odot}$. The lower panel shows the cumulative mass of stars formed versus time for clouds of $\sim 10^5$~M$_{\odot}$.}
\end{figure}

\begin{figure}
\includegraphics[scale=0.37]{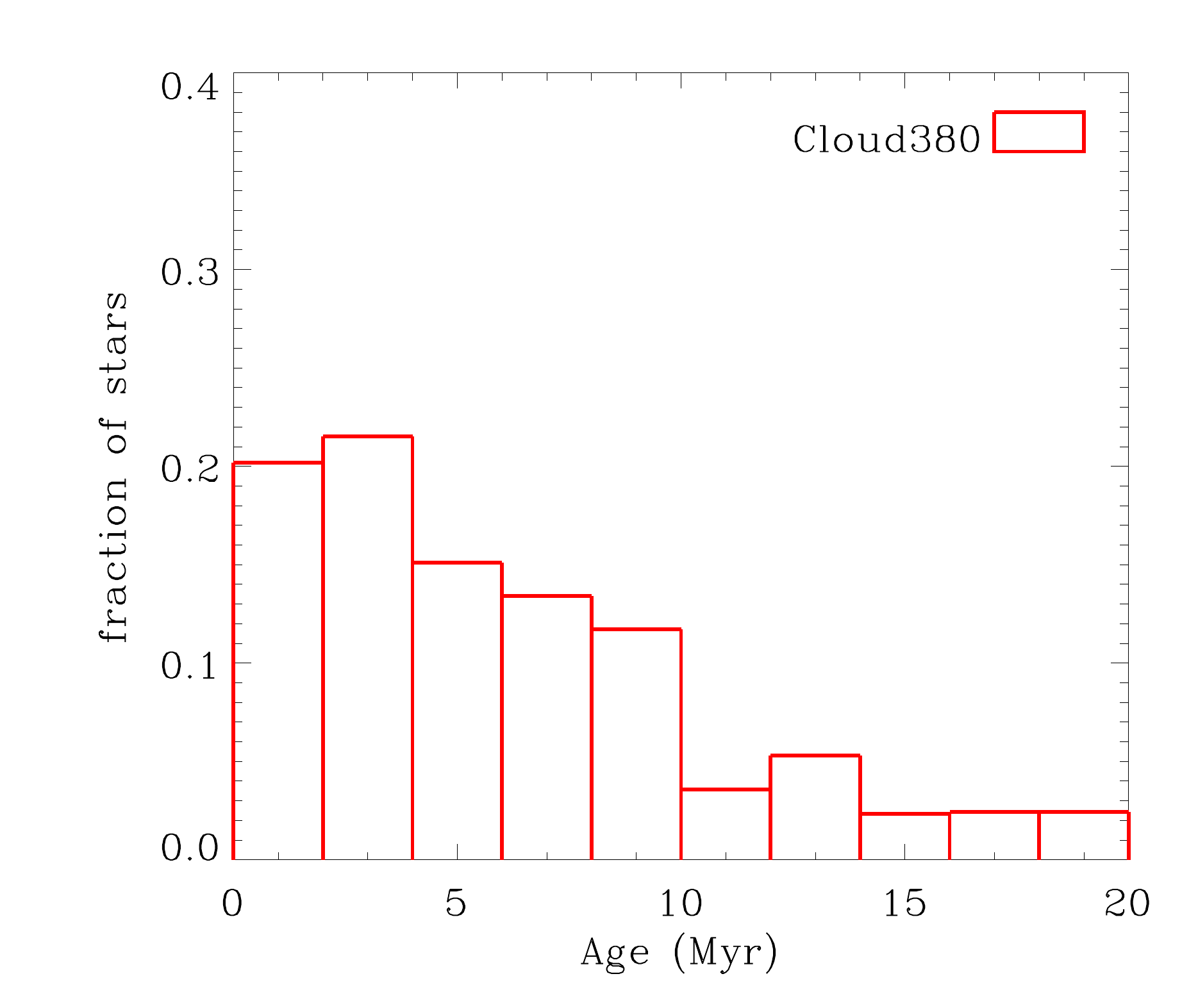}
\includegraphics[scale=0.37]{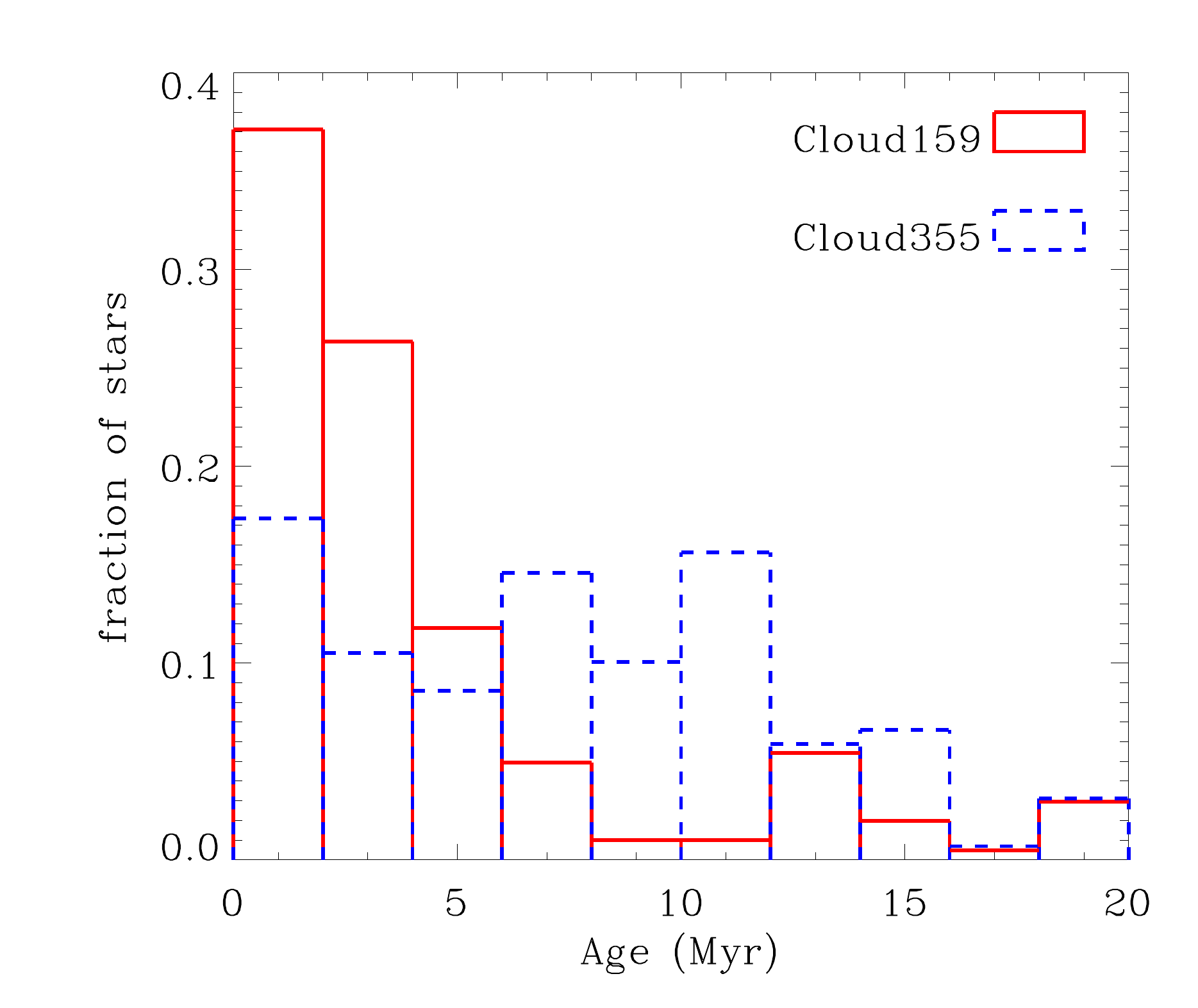}
\caption{The age distribution of stars which have formed from the constituent gas of Cloud380 (top) and Clouds 159 and 355 (lower) are shown. The distributions of ages are calculated at 250 Myr.}
\end{figure}

\subsection{Examples of other clouds}
So far we have concentrated on the evolution of a single cloud, Cloud380. However the behaviour of this cloud may not be typical behaviour for clouds in our simulation. In particular, Cloud380 is one of the most massive clouds in our simulation. Hence Cloud380 can only form by the agglomeration of smaller clouds (and ISM gas), and can only disperse by splitting into smaller clouds (and ISM gas). Thus Cloud380 is effectively at the top of the food chain of clouds in our simulation. Smaller clouds can form and disperse in a greater number of ways (a simple schematic is shown in Figure~10). Smaller clouds can form by the agglomeration of smaller clouds, \textit{and} form instantly as the byproduct of a dispersing larger cloud. Similarly smaller clouds can disperse due to feedback and/or shear, \textit{and} disappear instantly by becoming part of a larger cloud. Thus the `cloud lifetime', and evolution of the smaller clouds, can be somewhat different from both Cloud380 and usual evolution scenarios for molecular clouds (see e.g. \citealt{Vaz2010}). In this section we consider a number of other examples. 

In Table~1 we show the details of five other clouds we studied from the simulation. We tried to select clouds which contained at least a few 100 particles, and from a range of environments.  We include a cloud in the outer part of the disc, where the spiral arms are not very strong, and a cloud in the inter-arm region. Unfortunately the inter-arm cloud is not so well resolved as inter-arm clouds tend to be less massive (see also Columbo et al., in prep.). We do not show the detailed evolution of these clouds, but discuss briefly their evolution below. The only clouds where the gas does not demonstrably form from, or disperse into other smaller clouds, are those which are $<10^5$~M$_{\odot}$, which is not surprising as they are towards the resolution limit of the simulations, and may well undergo `artificial' formation or dispersal as mentioned in Section~4.3.2.
\begin{table*}
\centering
\begin{tabular}{c|c|c|c|c|c|c|c}
 \hline 
Cloud & Mass  & No. particles & $\alpha$ & Location at & Nature of cloud  & Nature of cloud \\
& ($10^5$ M$_{\odot}$) & & & $T_0=250$ Myr & evolution & dispersal \\
 \hline
 Cloud380 & 20 & 6386 & 2.9 & Spiral arm (R=3.1 kpc)& forms from and disperses into smaller clouds & shear + feedback\\
Cloud788 & 1.7 & 559 & 3.7 & Spiral arm  (R=4.3 kpc)& remains of, and progenitor of more massive cloud & feedback \\
Cloud877 & 3.1 & 999 & 1.8 & Spiral arm (R=4.1 kpc)&  forms from and disperses into smaller clouds & shear + unbound\\
Cloud355 & 0.96 & 305 & 3.6 & Inter-arm (R=3.3 kpc)&remains of more massive GMC & shear + unbound\\
Cloud159 & 2.7 & 863 & 2.7 & Outer disc (R= 8.3 kpc)& forms from and disperses into smaller clouds & unbound\\
Cloud1198 & 13 & 4291 & 0.8 &Spiral arm (R=3.4 kpc)& remains of more massive GMC & feedback \\
\hline
\end{tabular}
\caption{The different clouds examined in this paper. Outer disc refers to the outer region of the disc ($R>5$ kpc) where the spiral pattern is not particularly strong. The inner part of the disc is divided into spiral arm and inter-arm. Due to the complexity of determining cloud dispersal, the final column is somewhat speculative, but is based on Figure~5, $\alpha$, the shapes of the clouds, and number / distribution of nearby feedback events (see also text).}
\label{runs}
\end{table*}

\subsubsection{Cloud788}
Cloud788 is a $1.7 \times 10^5$ M$_{\odot}$ cloud in the spiral arm region, which over the course of 240-260 Myr, appears to emerge from the dispersal of a more massive ($4.3 \times 10^5$ M$_{\odot}$) cloud, and then part of Cloud788 goes on to form another ($2.5 \times 10^5$ M$_{\odot}$) cloud. This complex evolution is a consequence of Cloud788 lying in the spiral arm all this time, where there are merger events, and multiple feedback events. The mass of the `continuation of Cloud788' versus time is shown in Figure~6. The lifetime of this cloud is much less than Cloud380, being of order a few Myr.

\subsubsection{Cloud877}
Similar to Cloud380, Cloud877 is the highest point in food chain during its evolution. However there is only one 
feedback event during its evolution, suggesting it is shear as it leaves the spiral arm, and / or the unbound nature of the cloud, which is causing it to disperse.

\subsubsection{Cloud355}
Cloud355 is the remains of a more massive ($\sim 2.9 \times 10^5$ M$_{\odot}$) spiral arm cloud. Cloud355 is situated in a spur, and simply becomes more sheared out with time. There are a couple of feedback events between 250 and 260 Myr during the evolution of Cloud355, but probably the main mechanism of dispersal is shear. Both Cloud355 and Cloud877 are unbound, and less massive, so they can more quickly disperse.

\subsubsection{Cloud159}
Cloud159 appears to form and disperse in a similar manner to Cloud380, i.e. from the accumulation of smaller clouds, and dispersal into smaller clouds. However as Cloud159 lies at a much larger radius (8.3 kpc), shear is less likely to be able to disrupt the cloud (and would do so on a longer timescale, see Figure~5). In addition, there are no feedback events close to when the cloud disperses (unlike Cloud380), so feedback is unlikely to be responsible for cloud dispersal. Cloud159 appears relatively elongated ($\sim100$ pc $\times 20$ pc) and splits into two at the narrowest point of the cloud. In fact, Cloud159 forms from two clouds which adjoin, and then split apart again (consequently Cloud159 only has a very short life, of around 5 Myr). This suggests the larger scale gas dynamics are determining the evolution of Cloud159. The mass of the continuation of Cloud159 is shown in Figure~5. Similar to the other $10^5$ M$_{\odot}$ clouds, the lifetime is relatively short. The mass reduces less steeply with time, possibly because the cloud is subject to less shear. 

\subsubsection{Cloud1198}
Cloud1198 is the remains of a more massive ($\sim 2 \times 10^6$ M$_{\odot}$) cloud. Cloud1198 has the longest lifetime, exceeding the duration of the 40 Myr time period we consider (this is the only such example). Unlike Cloud380, even by 270 Myr Cloud1198 is not particularly elongated, and only one relatively small cloud has broken away, although the mass of Cloud1198 is decreasing. This suggests that feedback is slowly dispersing the cloud. This cloud is also the most bound of those we study individually, and presumably shear and large scale motions are not effective mechanisms for dispersing the cloud. 

\subsubsection{Ages and star formation of clouds}
In Figures 8 and 9 we show the star formation histories for some of these lower mass clouds. Figure~8 (lower panel) again shows a higher number of stars formed around 250 Myr. There is a peak in the star formation rate for between about 5 and 10 Myr, again suggesting that a `lifetime' of these clouds would be several Myr. However there are still a number of stars being formed after 255 Myr, showing that some of the gas is then accreting onto other clouds and continuing a small amount of star formation. The mass of stars formed from 245 to 255 Myr is $\sim$ 1500 M$_{\odot}$ for Cloud159 and Cloud788, which gives a relatively low efficiency of $\lesssim 1$ \%. 

Figure~9 (lower panel) shows the age distribution of stars in Cloud159 (the distribution is similar for Cloud788 and Cloud877).  The distribution is more strongly peaked at young ages, compared to the more massive Cloud380, and thus more closely resembles \citet{Hartmann2012}. Interestingly though, the distribution does not go to zero, and there are still a small number of stars with ages $>$ 10 Myr. For Cloud355 (Figure~9, lower), which represents the remains of a very massive GMC,  there is less evidence of a peak of very young stars. There is no particular decrease in the fraction of stars with age until around 12 Myr, after which there are considerably fewer stars. We caution again that as we do not include any star particles, we cannot say anything about cluster dispersal. Nevertheless our results suggest intriguing differences in the stellar population of GMCs depending both on their mass, cloud history and position in a galaxy.

\begin{figure}
\centering
\includegraphics[scale=0.45]{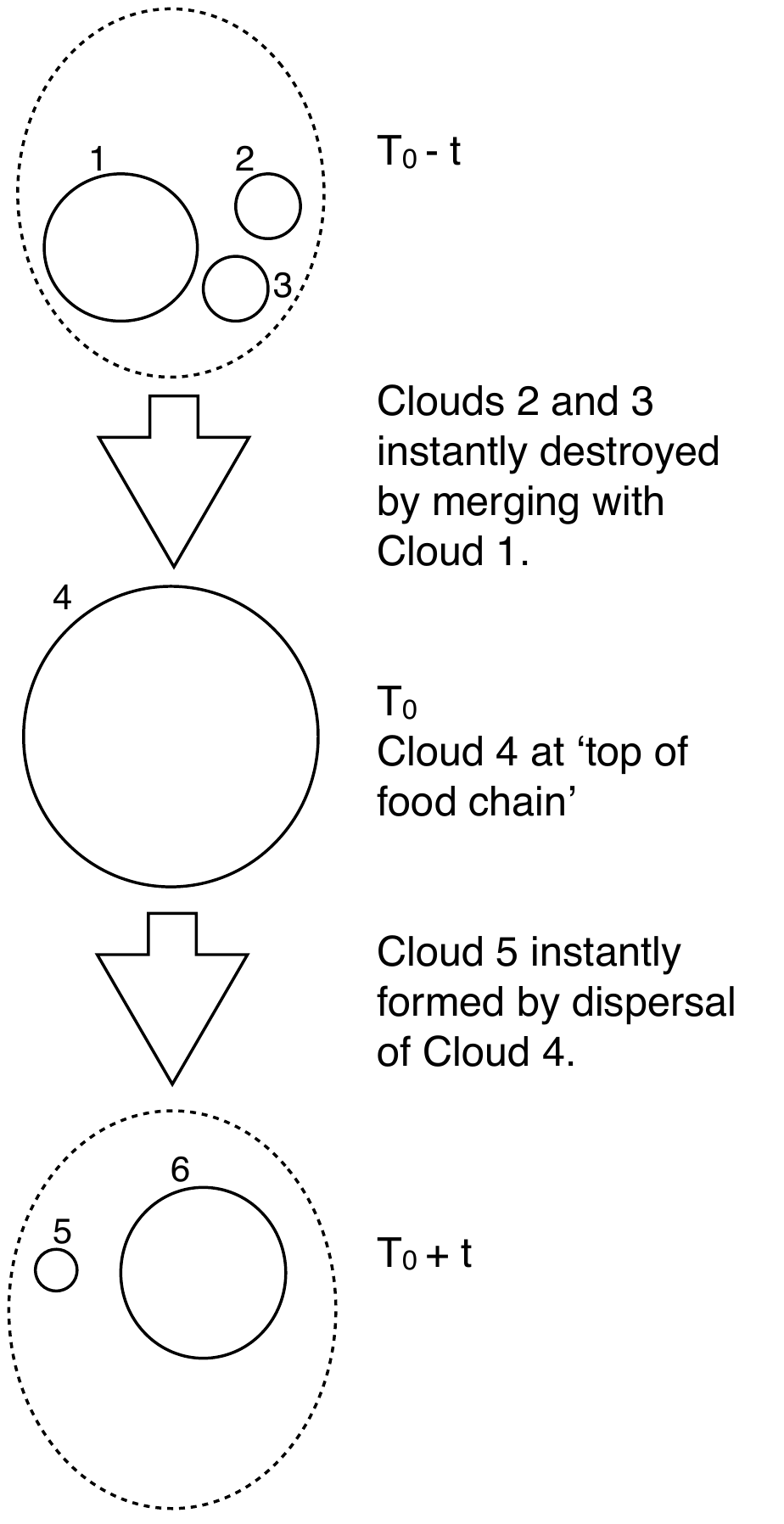}
\caption{A simple schematic of possible cloud evolution is shown. For simplicity clouds are represented by solid circles, and regions of ambient ISM by dashed ellipses. The evolution of Cloud 4, which forms from and disperses into a number of smaller clouds (like Cloud380), is perhaps closest to the typical picture of GMC evolution. However the evolution of smaller clouds can differ, forming from the dispersal of Cloud 4, or ending their lives as a new piece of Cloud 4 (see also Figure~1, \citealt{Tasker2009}).}
\end{figure}

\subsection{Statistical properties of all $>10^5$ M$_{\odot}$ GMCs}
So far we have concentrated on individual examples of clouds. Here we extend our analysis to consider collectively all $>10^5$ M$_{\odot}$ GMCs  in our simulation (a total of 61 GMCs).

\subsubsection{Cloud lifetimes}
We use the same method to calculate the lifetimes as used for Cloud380 in Section~5.1.5, but it is stated here more precisely. We compute the lifetime of each cloud as follows:
\begin{enumerate}
\item We follow the continuation of a given cloud by finding the constituent precursor or successor cloud at each time frame.
\item We compute the mass of original cloud particles in each precursor or successor of our chosen cloud.
\item We define the beginning of the cloud's life as the first point before $T_0$ when at least half of the original cloud particles are situated in a precursor cloud. 
\item We define the end of the cloud's life as the last point after $T_0$ when there are at least half of the original cloud particles in a successor cloud. 
\end{enumerate}
The time resolution for determining a cloud's lifetime is 1 Myr between 245 and 255 Myr, and 5 Myr otherwise.

In Figure~11, we show our measure of the GMC lifetime versus mass (blue points). The GMCs tend to have relatively short lifetimes, the most frequent around 5 Myr, though with a large spread from 2 to 40 Myr. The clouds with the longer lifetimes tend to be those which are more massive. The lower limit of 2 Myr is a genuine lower limit on the lifetimes of the clouds (the minimum we can resolve is 1 Myr). Clouds can have quite short lifetimes in our models because, for example, a cloud that splits into 2 equal parts will instantly `die' in our model. 

We also show in Figure~11 points where a lower threshold surface density (50 M$_{\odot}$ pc$^{-2}$) is used to find the clumps. The distribution of lifetimes is not dissimilar, though there are relatively more 10-25 Myr lifetime clouds and fewer clouds with lifetimes $\leq 5$ Myr. Equivalent, more massive, clumps found with the lower density threshold tend to have longer lifetimes (by a factor of around 2) compared to their counterpart clumps found with $\Sigma_{th}=100$ M$_{\odot}$ pc$^{-2}$, suggesting unsurprisingly that GMAs will likely have longer lifetimes than GMCs .

\begin{figure}
\centering
\includegraphics[scale=0.28]{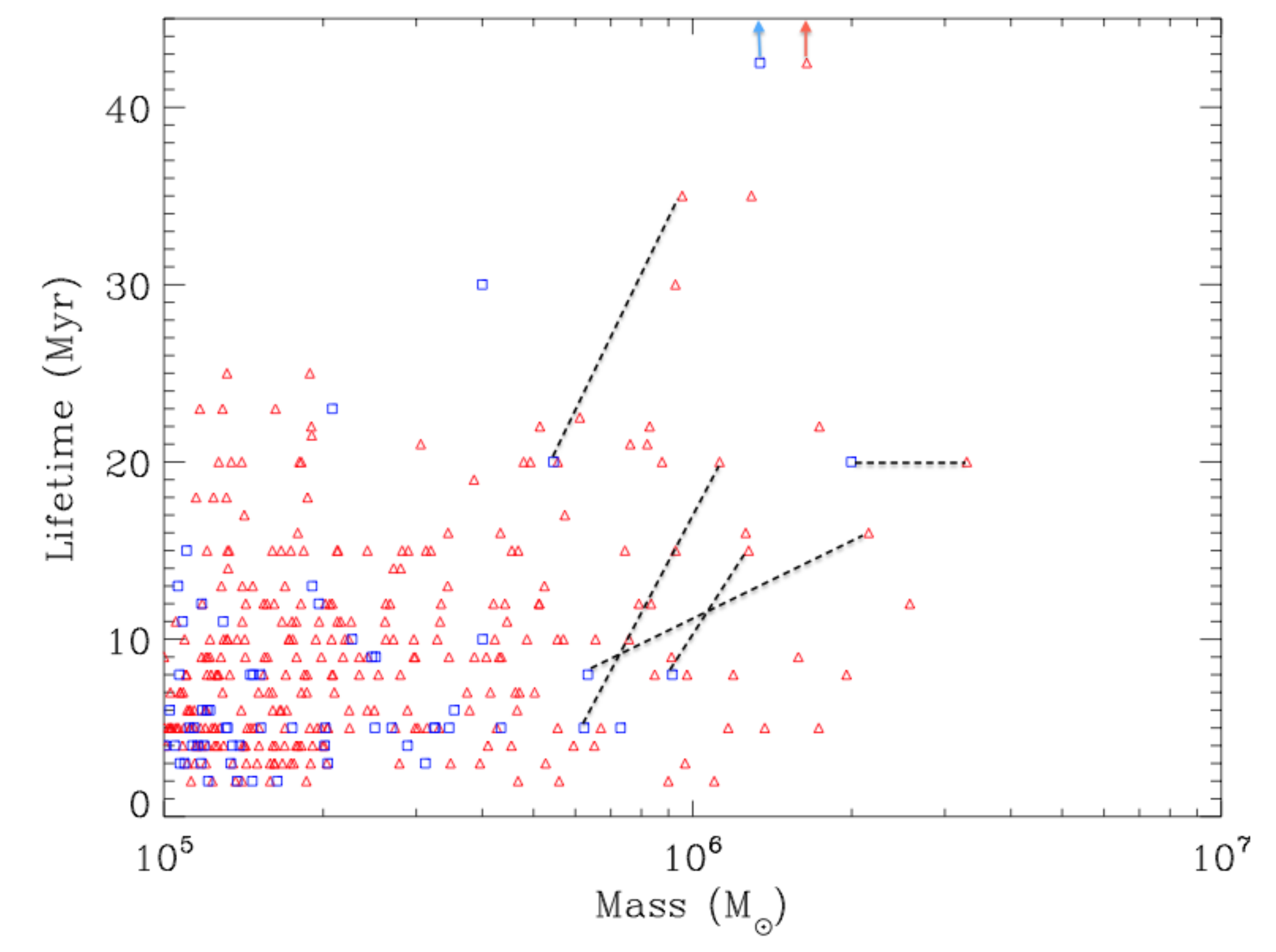}
\caption{The lifetime of clouds $>10^5$ M$_{\odot}$ is shown versus their mass. Most cloud lifetimes are between 2 and 25 Myr. The clouds with the longest lifetimes ($>$ 25 Myr) tend to be the most massive $>10^6$ M$_{\odot}$ GMCs. Points are shown when applying a critical surface density threshold of 100 (blue squares) and 50 (red triangles)  M$_{\odot}$ pc$^{-2}$ for the clump finding algorithm. For a few of the clouds, a dashed line connects equivalent clouds found with the different surface density criteria. The arrows represent lower limits on the lifetime of a cloud which exists for longer than the 40 Myr we consider.}
\end{figure}

\subsubsection{Cloud lifetimes and crossing times}
A cloud's lifetime is likely to be related to the velocity dispersion of the cloud, and its size, so essentially the crossing time of a cloud. \citet{Elmegreen2000} suggests that star formation and cloud dispersal occurs within 1 or 2 crossing times of a cloud. 
Of the clouds we considered, Cloud159 undergoes the most straightforward `termination', simply splitting in half at the narrowest point of the cloud (see Figure~12). The time for this cloud to disperse is then likely to be the length $L$ divided by the relative velocity of the sections either side of the part marked $L$ on Figure~12. Taking the velocity dispersion of the cloud, $\sigma=5$ km s$^{-1}$, as a proxy for the latter, we can estimate a timescale of $t=L/\sigma$. Taking $L=10$~pc we obtain a time of $t=$2~Myr. This is a reasonable estimate of the timescale to split up the cloud.
\begin{figure}
\centering
\includegraphics[scale=0.3]{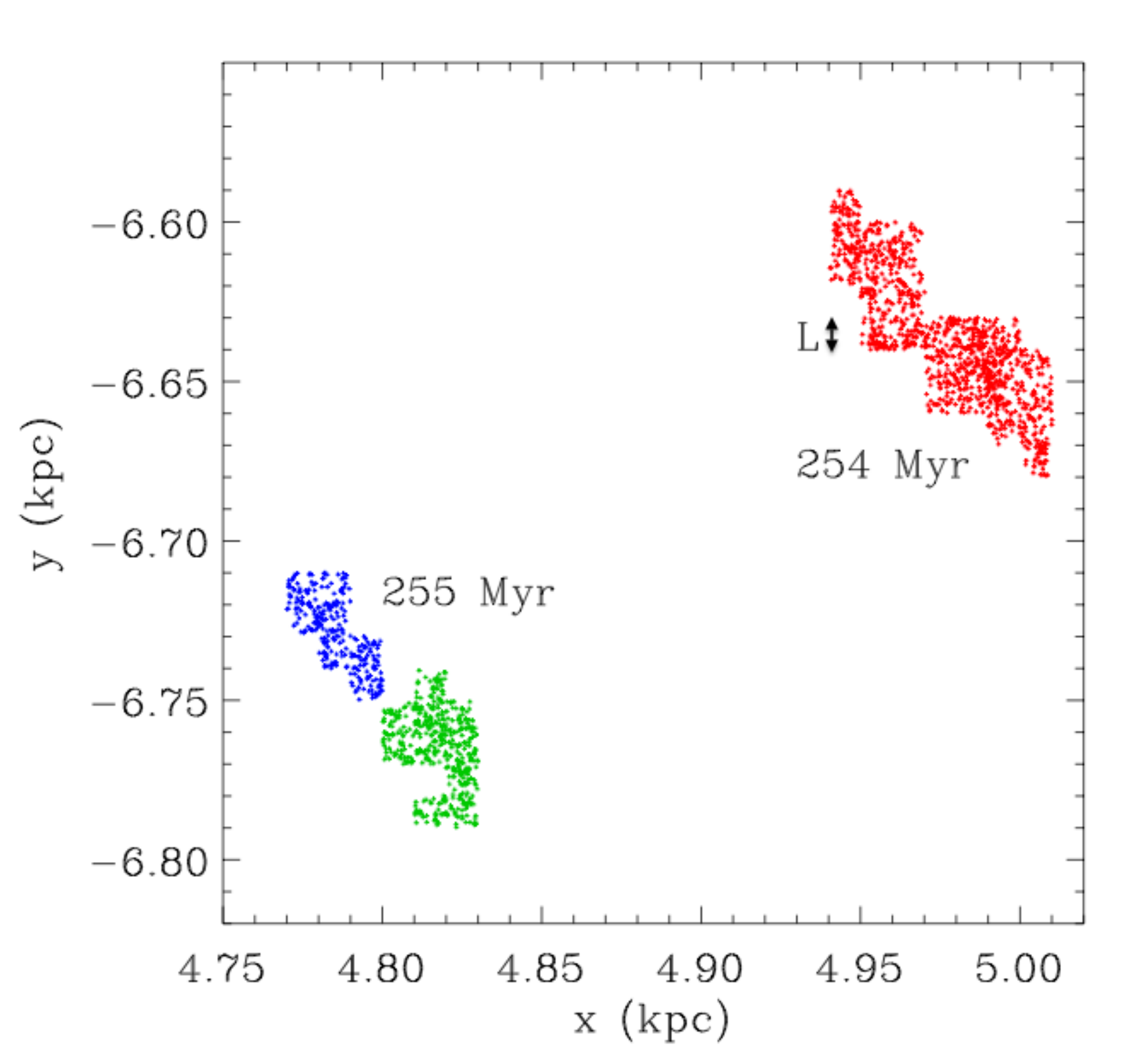}
\caption{The point at which Cloud159 disperses: at 254 Myr, the successor of Cloud159 is a single cloud, however at 255 Myr, this cloud has split into two. The times are only given to the nearest Myr, but the time difference between the two stages is 1.0 Myr.}
\end{figure}

\begin{figure}
\centering
\includegraphics[scale=0.43]{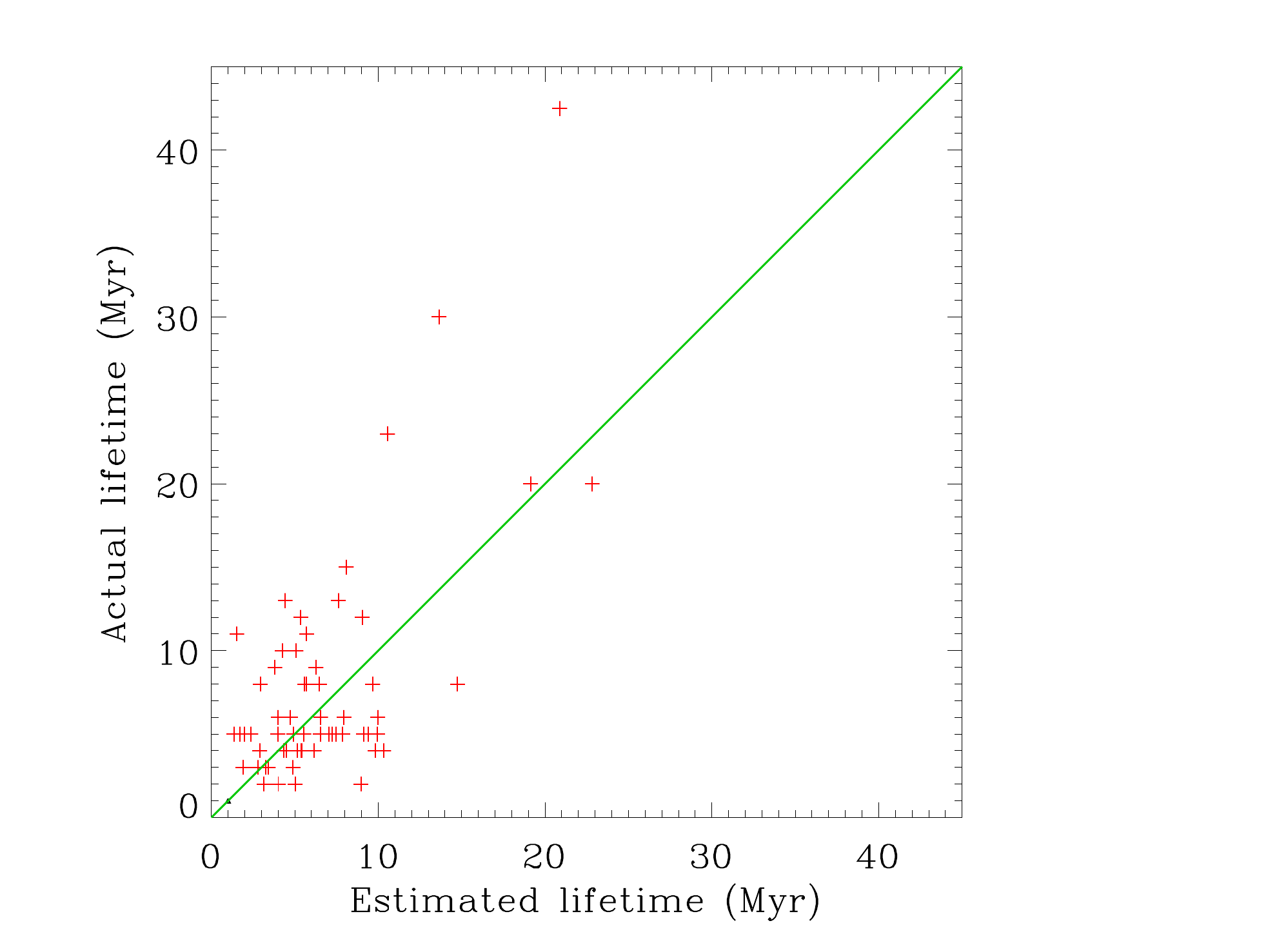}
\caption{The actual lifetime of clouds (defined in Section 5.3.1) is plotted against our estimated time given by Equation~4. Outlying clouds are often those with particularly elongated, or compact shapes.}
\end{figure}

Obviously we cannot study every cloud individually, to determine how, and where each cloud splits into multiple pieces. Furthermore larger clouds such as Cloud380 split up rather by many smaller clouds breaking away over a period of time. However we can expect that the time for clouds to disperse will depend on their dimensions, and their velocity dispersion. We can obtain a simple estimate of a cloud's age from the crossing time of a cloud.
We estimate the lifetime according to
\begin{equation}
t_{est}=\frac{2b}{\sigma}
\end{equation} 
where $b$ is the semi-minor axis of the cloud, and $\sigma$ the velocity dispersion of the cloud. Overall, feedback, shear and spiral shocks are likely to contribute to $\sigma$, as discussed earlier, though Equation~4 makes no attempt to incorporate these explicitly.
We include the factor 2 because when we select a cloud, it is not at the start of its lifetime. We make the simplest assumption that the cloud life is symmetric about the point which we observe it. 

In Figure~13 we show the measured lifetime of the GMCs versus this estimated lifetime.
There is a correlation between the actual and estimated lifetimes (the actual lifetime typically within a factor of 2 of the estimated lifetime), suggesting that to first order, the lifetime is related to the crossing time, however there is quite a lot of scatter.  Some scatter may simply be because we assumed the evolution of the cloud is symmetric about $T_0=250$ Myr, which is unlikely to be true for all clouds (see Figure~6). We also investigated whether the scatter was related to the aspect ratio, cloud porosity, virial parameter or star formation activity and examined individual outliers in Figure~13. We found that both the aspect ratio, and the cloud porosity contributed to the scatter, and outliers in Figure~13 were typically either very long, filamentary clouds or compact clouds. Thus although we take into account the shape of the cloud by using the semi-minor axis in Equation~4, this does not satisfactorily represent the complex shape of the cloud, or likewise the porosity. Therefore the shape of the cloud appears have a complex effect on its lifetime, which is difficult to disentangle. 
We also found a weak correlation of lifetimes with galactocentric radius. The lifetimes were found to increase slightly with radius, which we interpret as being due to the reduced role of shear at larger radii.
\begin{table*}
\centering
\begin{tabular}{c|c|c|c}
 \hline 
Fraction of clouds at  & Fraction of clouds previously  & Fraction of clouds which will form & Fraction of clouds which are at \\
 least 1/2 their maximum mass at $T_0$ & part of more massive clouds & more massive clouds in the future &  at a local minimum in mass \\
 \hline
64 \% & 15 \% & 16 \% & 5 \% \\
\hline
\end{tabular}
\caption{The percentage of clouds which are at their maximum mass, have previously been more massive, or are yet to reach their maximum mass are shown. The last column displays the fraction which were previously more massive, and will again be more massive GMCs within 20 Myr. For our purposes, we required that the cloud mass had to increase by at least a factor of 2 to avoid being placed in the first column.}
\label{runs}
\end{table*}

\subsubsection{Were (will) most clouds (be) constituent pieces of larger GMCs, or are they at the top of the food chain?}
In Section 5.2 we highlighted examples of some $\sim10^5$ M$_{\odot}$ clouds that were previously part of larger GMCs. Here we check whether GMCs are typically at their most massive when we observe them or not. We define 4 types of cloud:
\begin{enumerate} 
\item Clouds which are their most massive at $T_0$.
\item Clouds which were part of more massive GMCs prior to $T_0$.
\item Clouds which will become part of more massive GMCs after $T_0$.
\item Clouds which were part of more massive GMCs prior to $T_0$, \textit{and} become more massive GMCs again after $T_0$.
\end{enumerate}
We again take our sample of $>10^5$ M$_{\odot}$ GMCs at $T_0=250$ Myr and look for precursor or successor clouds between $T_0-20$ Myr and $T_0+20$ Myr which contain the same particles. For a more massive precursor or successor cloud to be classed as significant, we denote that it has to contain at least twice as much mass as the original cloud. Otherwise, the cloud is assigned to the first category. In Table~2 we show the percentage of different types of cloud. Nearly 2/3 of clouds are at least half of their maximum mass at the time they are selected ($T_0$). 30 \% of clouds were previously, or will be, part of GMCs which are twice as massive. The remaining few are clouds which were previously part of more massive GMCs and will again be part of massive GMCs. Thus most GMCs we observe are roughly at their maximum mass. However the fraction which become, or were previously, part of significantly bigger objects is still clearly important.

\section{Conclusions}
In this paper we have followed the detailed evolution of individual GMCs in a global galactic disc simulation.
We studied one $2 \times 10^6$ M$_{\odot}$ GMC, Cloud380, in particular detail. We showed that this GMC was formed from a mixture of clouds (mostly lying in the spiral arms), and ambient ISM (mostly inter arm). This particular GMC is effectively at the top of the food chain in terms of the evolution of molecular clouds and GMCs. 
Cloud380 likewise disperses into smaller clouds (lying in a `spur' feature) and ambient inter arm gas. The tendency of the most massive GMCs to evolve into spurs containing GMCs, \textit{before} more massive clouds start to build up in the arm again, can account for recent observations that in some sections of spiral arm, the most massive clouds, and the most star formation, occurs in spurs \citep{Schinnerer2013}. 

More generally, other smaller clouds can have different evolutionary scenarios. Clouds can be immediately created by the dispersal (or splitting up) of another cloud. Clouds can also immediately lose their identity by becoming part of another cloud. Thus the history for different clouds will likely be quite varied.

Because of this complex evolution of GMCs, that the precursor or successor of a cloud may not be obvious, and that constituent material in clouds changes with time, assigning a lifetime to clouds is difficult. Here we assign a lifetime as the time over which, if we follow the evolution of a given cloud, its mass does not decrease by more than half. We further stipulated that this mass had to be the same constituent material as when the cloud was selected. We find that the minimum cloud lifetime is 2 Myr, but there is a large range in cloud lifetimes. Most lifetimes are between $\sim4$ and 25 Myr. An alternative lifetime could be based on the length of time gas remains molecular, though this would mean molecule rich galaxies would have long-lived clouds, regardless of whether the constituent gas of those clouds remains the same with time.

We attempted to predict the lifetime of clouds in our simulation, by computing their crossing time. Although this can provide a rough estimate, the complexities of cloud shapes and dynamics leads to a lot of scatter, when plotted agains their true lifetimes. Clouds which are very filamentary can easily be broken apart, whereas those which have small aspect ratios, and little porosity cannot be readily split in half. Furthermore as the lifetimes of more massive clouds are longer, the velocity field may well change on the timescale of the cloud's evolution. We have here looked at shear and feedback as possible mechanisms for dispersing clouds. Simply the morphology of spurs in galaxies is sufficient to suggest that shear is important. We suggest that shear may be responsible at smaller galactic radii, and on larger (tens of pc) scales, whereas feedback is more important on smaller scales. 

A caveat to these results is that we included instantaneous feedback. In some cases, the lifetimes and dynamics of clouds may be governed by the larger scale velocity dispersion of the gas, and be less dependent on local feedback events. However we leave it to future simulations and tests to fully examine the effect of different feedback prescriptions. We also plan to study the GMCs in terms of CO and HI emission, rather than total density, in a future paper.

We also examined the star formation history of molecular clouds, although we did not include star particles to follow the evolution of stars. Nevertheless, we found that the age distributions of stars are generally heavily skewed towards younger stars, but there is some variation according to the history of the cloud (e.g. whether it has formed from the dispersal of a more massive cloud).
We also showed that over the lifetimes estimated for our clouds, the constituent gas undergoes much more star formation, as expected, and the star formation efficiencies of the clouds (as measured by the mass of stars formed divided by the mass of the clouds) are of of order 1 \%. In future work we will include star particles to investigate star formation in clouds more consistently. 

Finally, we have also examined the properties of GMCs formed in our simulation (Appendix). The properties of the GMCs (mass function, rotation, virial parameter) are similar to those in previous simulations \citep{Dobbs2011old,Tasker2009,Dobbs2008}. We do not find much dependence of GMC properties on the criteria used for our clump--finding algorithm, or the resolution of the simulation.

\section{Acknowledgments}
We would like to thank an anonymous referee for a helpful report.
The calculations for this paper were performed on the DiRAC Facility jointly funded by STFC, the Large Facilities Capital Fund of BIS, and the University of Exeter. 
CLD acknowledges funding from the European Research Council for the 
FP7 ERC starting grant project LOCALSTAR. Figures included in this paper were produced using \textsc{splash} \citep{splash2007}.
\bibliographystyle{mn2e}
\bibliography{Dobbs}

\appendix 
\section{Properties of GMCs}
\begin{figure}
\centerline{
\includegraphics[scale=0.45]{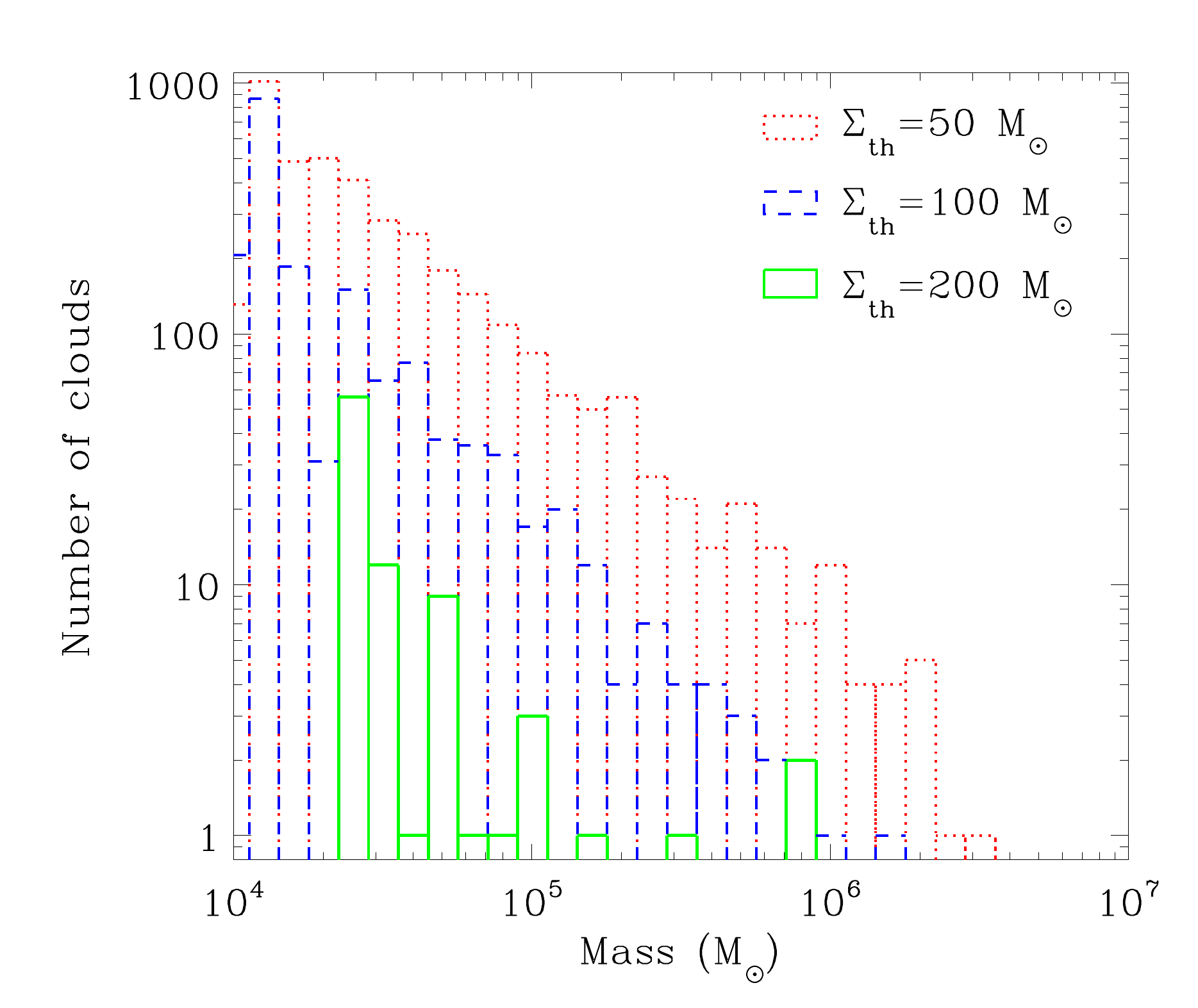}}
\centerline{
\includegraphics[scale=0.44]{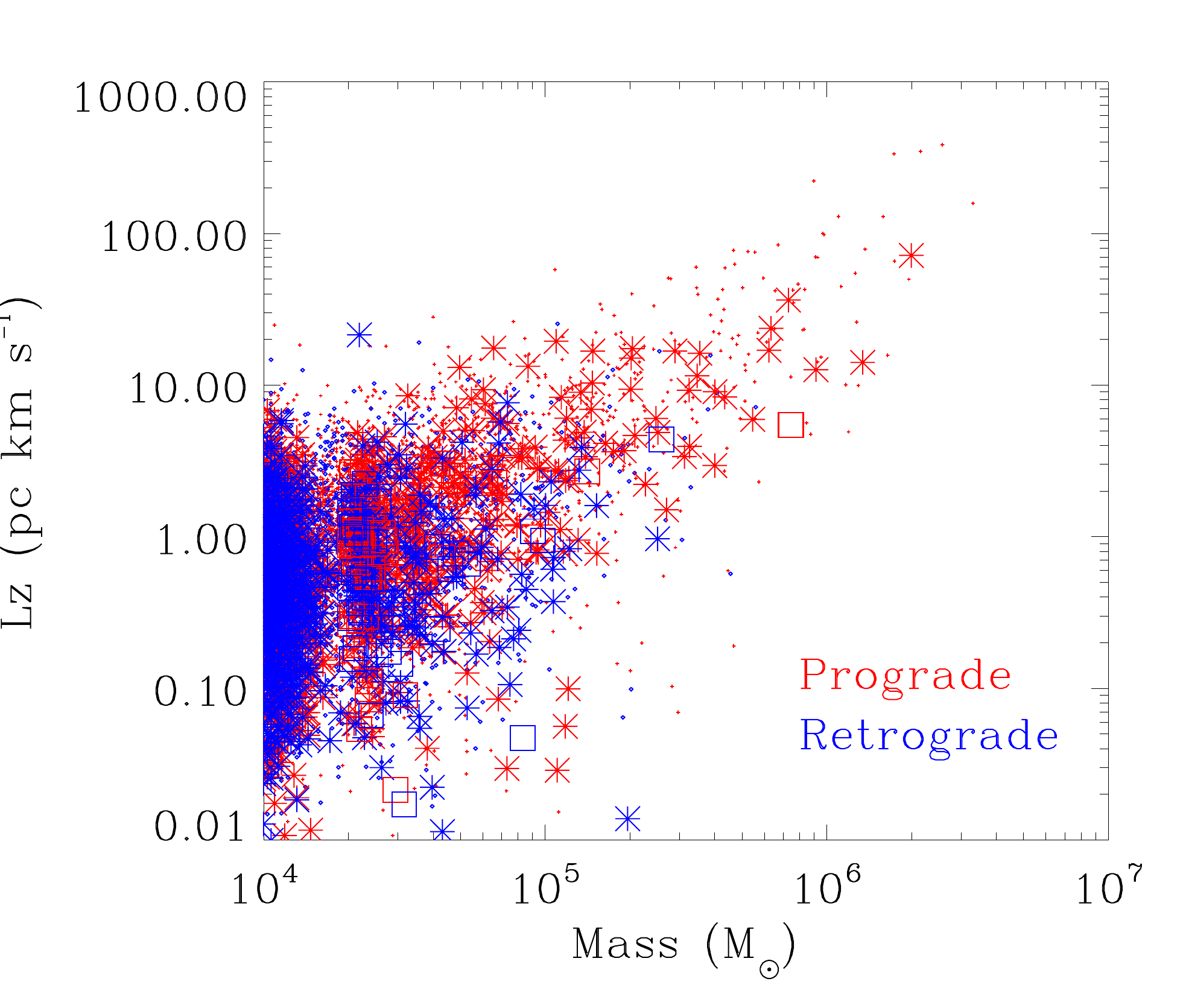}}
\centerline{
\includegraphics[scale=0.42]{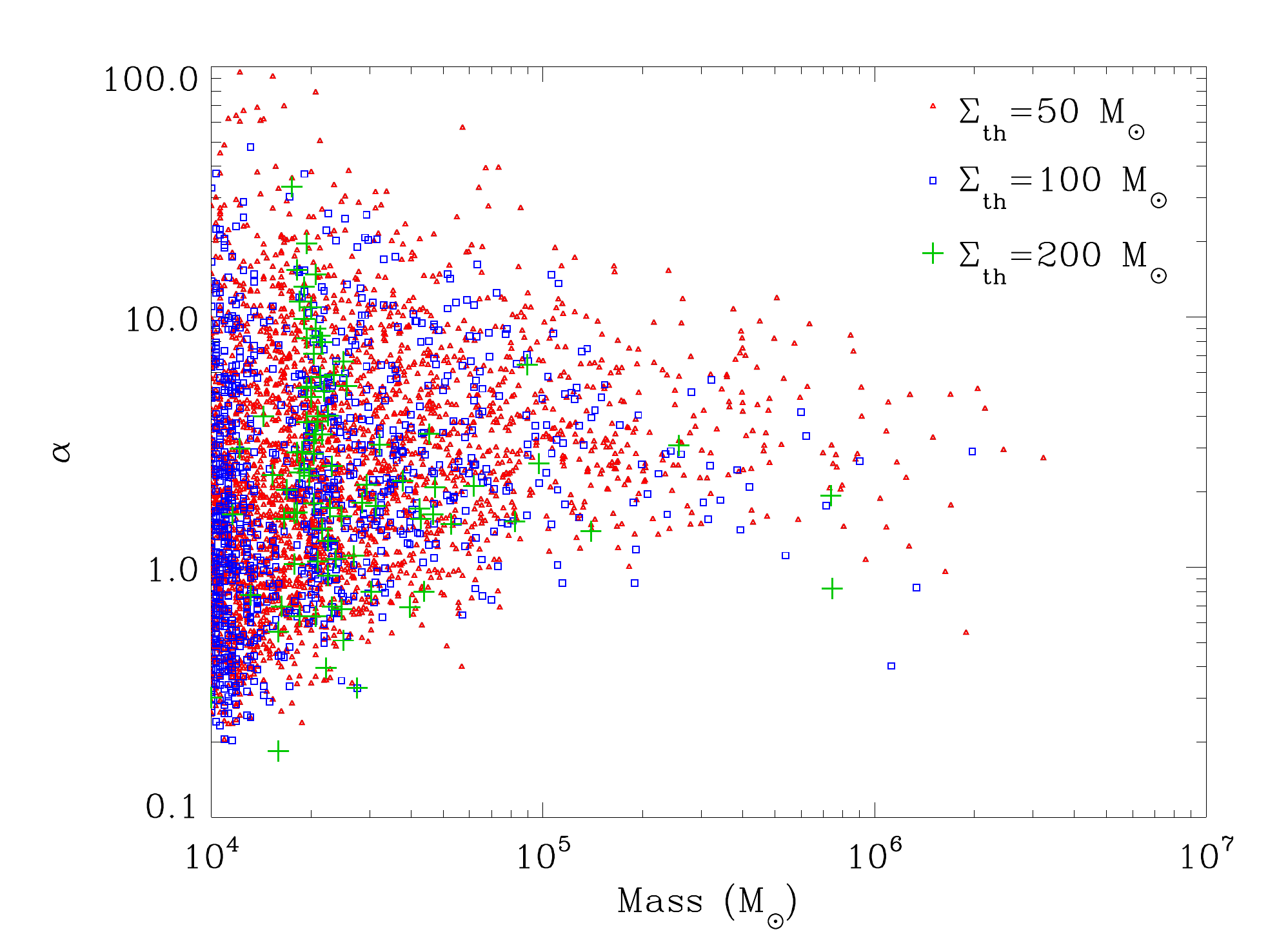}}
\caption{The cloud mass function (top), cloud rotation (centre) and virial parameters (lower) are shown for clouds selected from our the simulation presented in this paper. We tried three different column density thresholds to extract the clouds, $\Sigma_{th}$=50, 100 and 200 $M_{\odot}$ pc$^{-2}$. In the middle panel, the symbols represent $\Sigma_{th}$ =50 M$_{\odot}$ pc$^{-2}$ (points), $\Sigma_{th}$ =100 M$_{\odot}$ pc$^{-2}$ (asterisks) and $\Sigma_{th}$ =200 M$_{\odot}$ pc$^{-2}$ (boxes). For $\Sigma_{th}$ =200 M$_{\odot}$ pc$^{-2}$, there are no low mass clouds. This is because we would require at least  $2\times10^4$~M$_{\odot}$ in a 10 pc $\times$ 10 pc cell (used for our clump finding algorithm) to reach a surface density of  $200$~M$_{\odot}$ pc$^{-2}$.}
\end{figure}
In Figure~A1 we show the properties of GMCs found in our simulation. Compared to our previous results, \citep{Dobbs2011new}, we now extend properties down to $10^4$ M$_{\odot}$. We show results for three different  implementations of our clump finding algorithm, where we take the minimum surface density required to define clumps, $\Sigma_{th}$, as 50, 100 and 200 M$_{\odot}$ pc$^{-2}$.  The properties of the clouds as shown in Figure~A1 appear relatively similar regardless of the column density threshold. The slope of the mass spectrum appears to become slightly steeper as the threshold increases, the spectrum at $\Sigma_{th}=50$ M$_{\odot}$ pc$^{-2}$ being rather flatter. With $\Sigma_{th}=200$ M$_{\odot}$ pc$^{-2}$ we obtain relatively few clouds, and a less coherent spectrum with a few outliers at high masses. There are no low mass clouds with $\Sigma_{th}=200$ M$_{\odot}$ pc$^{-2}$ because at least  $2\times10^4$~M$_{\odot}$ is required in a 10 pc $\times$ 10 pc cell (used for our clump-finding algorithm) to reach a surface density of  $200$~M$_{\odot}$ pc$^{-2}$.

The rotation of the clouds is also similar regardless of the threshold for the clump-finding algorithm. The main difference is simply the cut off in mass, though the values of angular momenta also appear slightly lower with $\Sigma_{th}=200$ M$_{\odot}$ pc$^{-2}$. The values of $Lz$, and relative fractions of retrograde clouds are similar to our previous results \citep{Dobbs2008,Dobbs2011new}. The fraction which are retrograde are 39~\% ($\Sigma_{th}$ =50 M$_{\odot}$ pc$^{-2}$), 40~\% ($\Sigma_{th}$ =100 M$_{\odot}$ pc$^{-2}$) and 44~\% ($\Sigma_{th}$ =200 M$_{\odot}$ pc$^{-2}$). The most massive clouds (for any particular $\Sigma_{th}$) tend not to be retrograde though.

The virial parameters of the cloud are shown in the third panel of Figure~A1. As found in \citet{Dobbs2011old}, the majority of clouds tend to be unbound. The proportion of bound ($\alpha<1$) clouds increases from 10\% at the lowest threshold to 20\% at the highest $\Sigma_{th}$, which is not surprising as we would expect more dense clouds to be bound.

The number of clouds does obviously differ with the surface density threshold chosen. For the simulations here, this simply reflects the higher fraction of the gas which lies above the lower threshold.

\subsection{Resolution test}
We also performed a simulation with 1 million particles to see how the structure of the disc and properties of clouds change with resolution. Due to the non-trivial dependence of our feedback prescription on resolution, it is not possible to run an exactly equivalent simulation. However, as shown by Figure~A2, the structure of the disc, and appearance of inter-arm spurs are similar at the different resolutions, albeit there is less dense gas, and less obvious dense GMCs at the lower resolution. 

\begin{figure}
\centerline{
\includegraphics[scale=0.3]{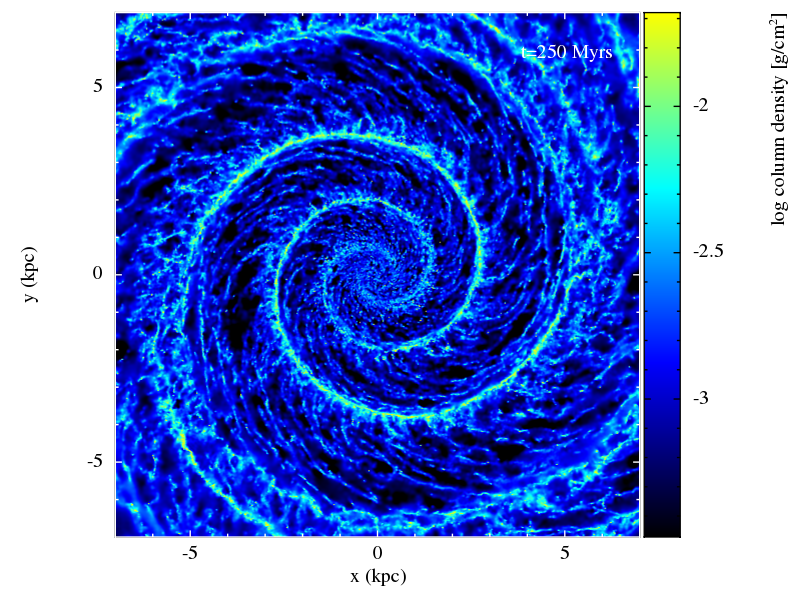}}
\caption{The column density for the galaxy disc is shown for a lower resolution simulation with 1 million particles.}
\end{figure}

\begin{figure}
\centerline{
\includegraphics[scale=0.45]{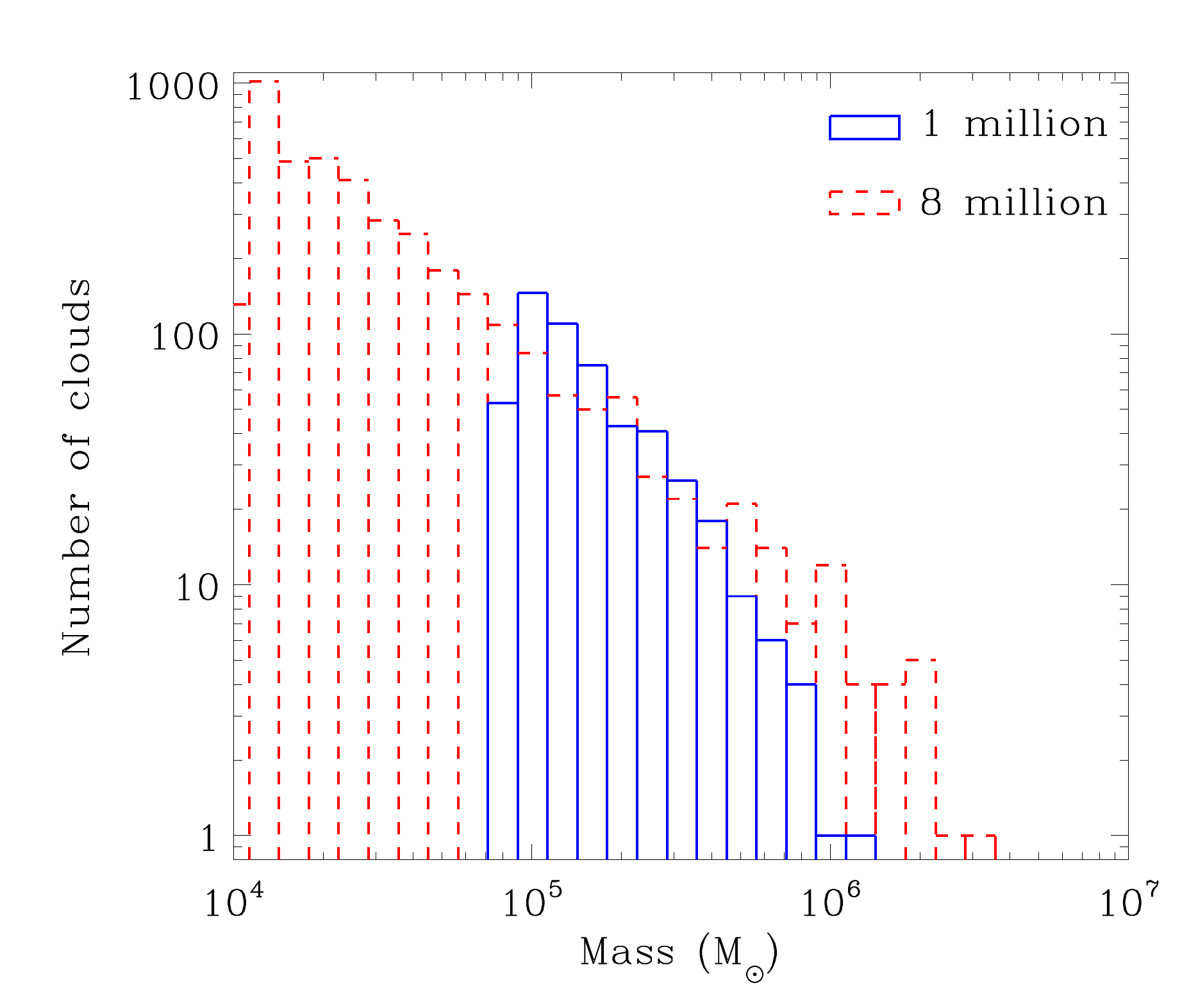}}
\centerline{
\includegraphics[scale=0.44]{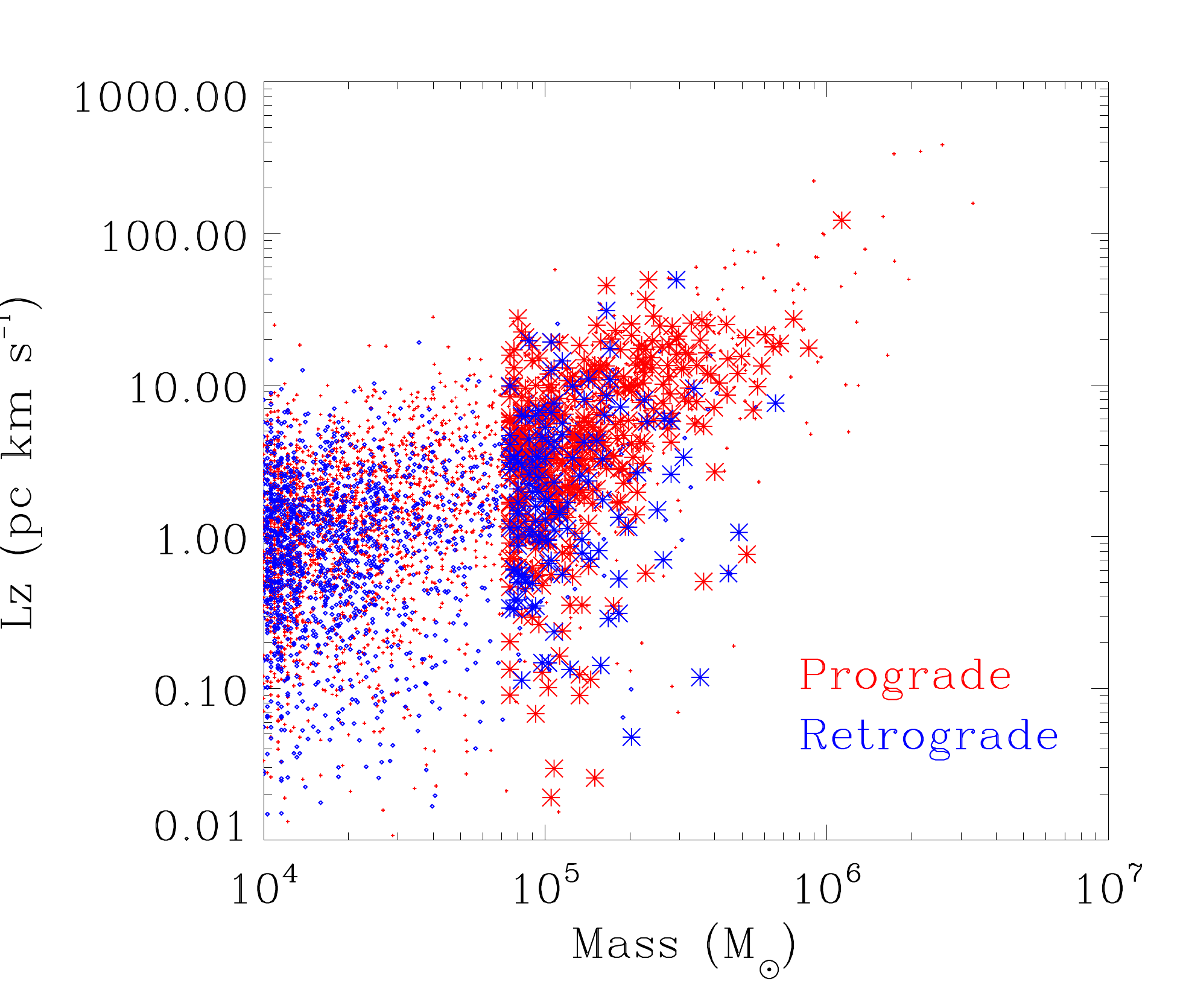}}
\centerline{
\includegraphics[scale=0.42]{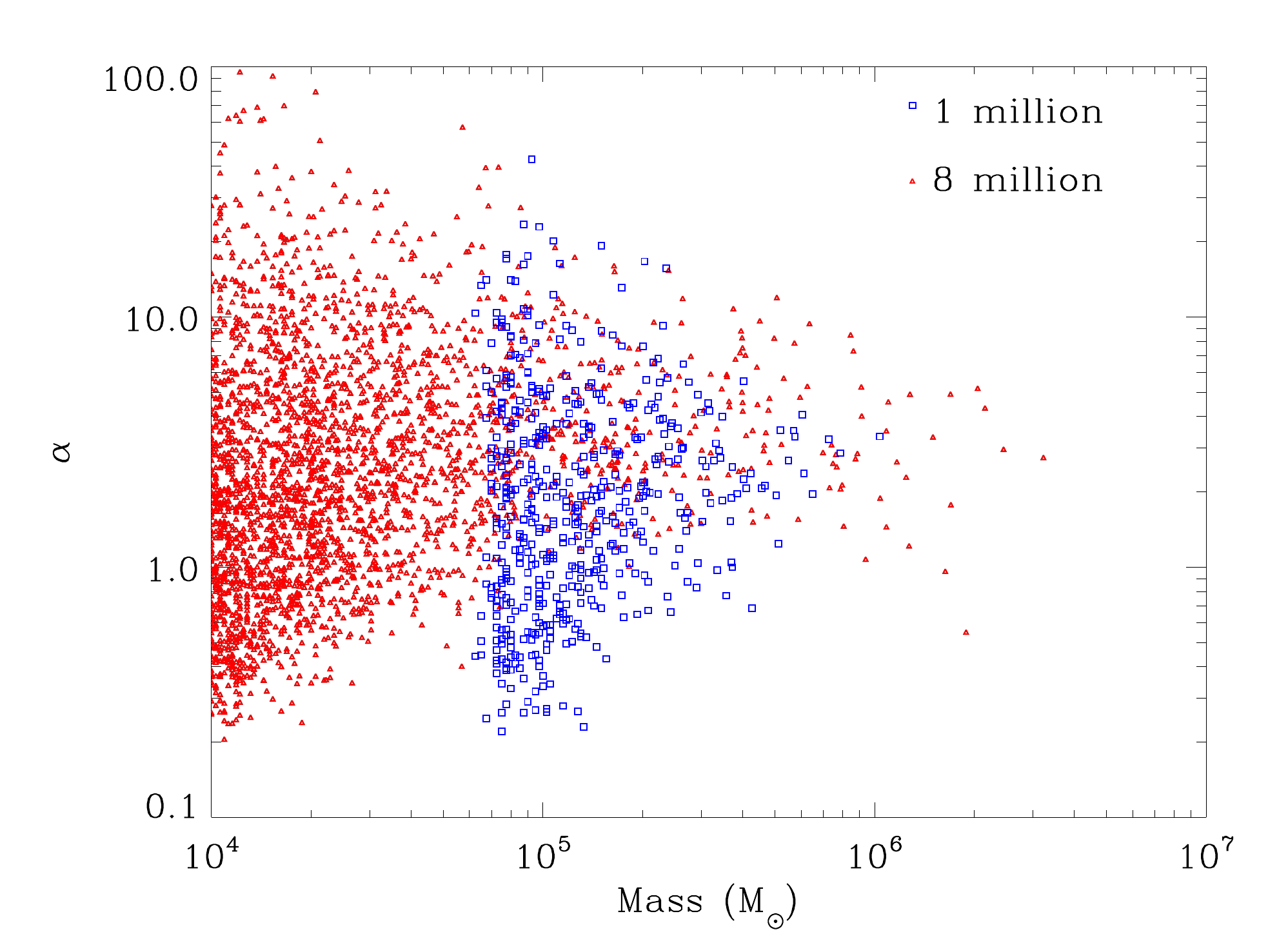}}
\caption{The cloud mass function (top), cloud rotation (centre) and virial parameters (lower) are shown for clouds 
from simulations with 1 and 8 million particles, using $\Sigma_{th}$=50 $M_{\odot}$ pc$^{-2}$ in our clump-finding algorithm. In the middle panel, the results for the 1 million particle simulation are represented by asterisks, and those for the 8 million particle simulation with dots.}
\end{figure}

We compare the properties of GMCs formed in the 8 and 1 million particle simulations in Figure~A3. In each case we take $\Sigma_{th}$~=50 M$_{\odot}$ pc$^{-2}$. We chose this value because taking a higher $\Sigma_{th}$ produced a too small number of clouds in the low resolution simulation to make meaningful comparisons. We see from Figure~A3 that there are some differences between the different resolutions, although we stress again that the simulations will not be exactly equivalent in the amount of feedback. The mass spectrum for the lower resolution simulation (top) is slightly steeper\footnote{The power spectrum also looks more like a curve than a power law. Departures from simple power laws, e.g. multiple slopes or truncated power laws are often found to be better fits for GMCs in nearby galaxies \citep{Hughes2013}.} This could be related to the finding in the previous section that for stricter criteria (relative to the resolution and properties of the simulation) the spectrum becomes steeper. It is unclear how significant these differences are though.

The distribution of the specific angular momenta of the clouds is very similar in the different resolution simulations. The only difference is that there is no obvious truncation in the mass of retrograde clouds in the lower resolution simulation compared to the higher resolution simulation (see also \citet{Dobbs2011new}), it is not entirely clear why this is. 

Finally the virial parameters of the clouds are shown in the lower parameter of Figure~A3. The range of $\alpha$ is similar for both resolutions. The clouds appear slightly more bound in the lower resolution simulation, probably indicating that the feedback is slightly stronger in the higher resolution simulation. The large spread in $\alpha$ for the lowest mass clouds appears to be an artefact (see also \citealt{Dobbs2011old}), though there always appears to be a decrease in the spread towards higher masses. The latter may well be a selection effect - those clouds which tend to reach the highest masses are those which are the most bound.

\bsp
\label{lastpage}
\end{document}